%% file: Paper.tex
\documentclass[twocolumn, 10pt]{IEEEtran}
\usepackage{hyperref} 
\usepackage{graphicx}
\usepackage{amssymb}
\usepackage{amsmath}
\usepackage{mathtools}
\usepackage{cite}
\usepackage{stfloats}
\usepackage{subfigure}
\usepackage{epstopdf}
\usepackage{psfrag}
\usepackage[mathscr]{euscript}
\usepackage{acronym}  
\usepackage{booktabs} 
\usepackage{cases}
\usepackage[table]{xcolor}
\usepackage[scaled]{helvet}
\usepackage{xparse}
\usepackage{bm}

\input{SupportDocuments/acronym}
\input{SupportDocuments/defmetric}

\begin{document}

\newcommand{\paperTitle}{
Securing Fresh Data in Wireless Monitoring Networks: Age-of-Information Sensitive Coverage Perspective
}
 

\title{\paperTitle}

\author{
	\vspace{0.2cm}
	Jinwoong~Kim, 
	Minsu~Kim, 
	and
	Jemin~Lee, \textit{Member, IEEE}
	%
	%
	%
	\thanks{
		J.\ Kim, M.\ Kim, and J.\ Lee are with the Department of Information and
		Communication Engineering, Daegu Gyeongbuk Institute of Science and
		Technology, Daegu 42988, South Korea
		(e-mail: \texttt{yoy876@dgist.ac.kr} \texttt{ads5577@dgist.ac.kr}, \texttt{jmnlee@dgist.ac.kr}).
	}
	\thanks{
		The material in this paper was presented, in part, at the Global Communications Conference, Taipei, Taiwan, Dec. 2020. \cite{KimKim20}
	}
\thanks{
	The corresponding author is J. Lee. 
}
}

\maketitle 
	\vspace{-20mm}

%

%

%
\setcounter{page}{1}
\acresetall

\acresetall

\begin{abstract}
With the development of IoT, the sensor usage has been elevated to a new level, 
and it becomes more crucial to maintain reliable sensor networks. 
%
In this paper, we provide how to efficiently and reliably manage the sensor monitoring system for securing fresh data at the data center (DC).
A sensor transmits its sensing information regularly to the DC, and the freshness of the information at the DC is characterized
by the age of information (AoI) that quantifies the timeliness of information. 
By considering the effect of the AoI and the spatial distance from the sensor on the information error at the DC, 
we newly define an error-tolerable sensing (ETS) coverage as the area that the estimated information is with smaller error than the target value. 
%
We then derive the average AoI and the AoI violation probability of the sensor monitoring system, 
and finally present the $\eta$-coverage probability, which is the probability that the ETS coverage is greater than $\eta$ ratio of the maximum sensor coverage. 
%
We also provide the optimal transmission power of the sensor, which minimizes the average energy consumption while guaranteeing certain level of the $\eta$-coverage probability. 
Numerical results validate the theoretical analysis and show the tendency of the optimal transmission power according to the maximum number of retransmissions. This paper can pave the way to efficient design of the AoI-sensitive sensor networks for IoT. 
\end{abstract}



\section{Introduction}\label{sec:models}

The \ac{IoT} and its enterprise-grade counterpart, the Industrial Internet of Things (IIoT)
have been giving significant and positive influence to our lives and industries.  
With the development of the \ac{IoT} technology, the sensor usage has been elevated to a new level. 
Sensors are devices that detect and respond to changes in an environment such as light, temperature, and motion.
The reliable communication of the sensors becomes more crucial, especially for time-critical applications such as fault detection and emergency alarms. 
Therefore, to support the various applications of IoT, 
the maintaining reliable sensor network has been one of the challenges in IoT. 
 

The design of reliable wireless sensor networks has been studied in many literature including the sensing coverage design\cite{YazSen14, ElhShi19, WanZhu16, YuaZha14}.
The sensing coverage means the area that a sensor can provide valid information from its measurement.
The disk-based coverage model has been used as a simplest model in \cite{YazSen14},
and Elfes sensing model, which stochastically covers partial areas, was proposed in \cite{ElhShi19}. 
The sensing coverage was also determined by exploiting the spatial correlation of sensed information in \cite{YuaZha14,WanZhu16}. 
However, most of works in sensing coverage design did not consider the freshness of sensed information although the data freshness can affect the accuracy or the value of the information, especially in realtime applications
such as health monitoring systems and road traffic reports \cite{Abd20}.



The freshness of the data has been studied intensively after the concept of \ac{AoI} has been proposed.
The \ac{AoI} is a novel metric for measuring the data freshness, defined as the time elapsed since the generation of the most recently updated information \cite{KauYat12}.
In the early studies of the AoI, many works focused on the theoretical analysis of the AoI 
by considering the various queueing models \cite{KauYat12, Mollei19, Bedsun19, CosCo16, DevDur18, SeoChoi19}. 
For instance, the average AoI of M/M/1, M/D/1 and D/M/1 queues is investigated under the first-come-first-served (FCFS) policy in \cite{KauYat12}.
The average AoI of multi-source M/G/1 queue with the FCFS policy is derived in \cite{Mollei19}, 
while the average AoI and average peak AoI are analyzed with multi-server queueing model based on the last-come-first-served (LCFS) policy in \cite{Bedsun19}
and with M/M/1/2* queue in \cite{CosCo16}.
The peak AoI outage probability has been presented for Geo/G/1 and D/G/1 queues, respectively, in \cite{DevDur18, SeoChoi19}. 
 Furthermore, unlike the general AoI model in which the age increases linearly with time, 
 a non-linear AoI model is also proposed in \cite{KosPa17, SunCyr19} to represent the level of discontent for data staleness over time.
 Recently, as a variation of the AoI, the age of synchronization (AoS) is also presented in \cite{TanWan19}
 to consider asynchronousness issues between the AoI at the receiver and that at the source.


For communication reliability in AoI-based systems, the retransmission techniques have also been considered\cite{DonLi20, BaoYu20, GuChen19, GonChe18}.
The average AoI is minimized by optimizing the redundancy allocation in the \ac{HARQ} scheme in \cite{DonLi20}. In \cite{BaoYu20}, the effect of the retransmission on AoI reduction is analyzed in the short packet-based machine type communications. In \cite{GuChen19}, the transmission power and the maximum allowable retransmission times are optimized to minimize the average AoI. {However, as the feedback is not considered, the unnecessary transmissions can occur even after the successful update.} 
The age-energy tradeoff is analyzed by comparing the retransmission and a transmission of newly generated update which pays for the sensing energy in \cite{GonChe18}. 

Recently, for efficient design of AoI-based systems, the correlation among generated information at different sources has been exploited\cite{PooBha17, Hricos19, SunCyr18}.
In \cite{PooBha17}, the average AoI was minimized by exploiting the temporal correlation between the consecutive updates. In \cite{Hricos19}, a correlated sensor apart from the exact source was exploited to decrease the sensing rate bearing a spatial error. The mutual information between the real-time source and received status updates are analyzed to quantify the freshness of information and optimized the sampling policy that maximizes the average mutual information in \cite{SunCyr18}. 
However, those works did not jointly consider the spatial and temporal correlation of sensing information, which fails to further enhance the sensing efficiency in the AoI-based system. 

{In this paper, we consider a sensor monitoring system, 
{where a sensor periodically transmits the sensed regional information to a \ac{DC} that requires maintaining fresh information, i.e., lower AoI. 
Within a sensing period, the retransmission scheme is also adopted with the limited maximum number of retransmissions. 
%
%
By exploiting the linear relationship between the AoI/the spatial distance from the sensor with the information error at the \ac{DC},
we define the \ac{ETS} coverage, defined as the area that the estimated information at the \ac{DC} is with smaller error than the target value.
%
Under such circumstances, we characterize the components (e.g., time interval of successful updates at the \ac{DC}) that affects the \ac{AoI}, and derive the average AoI of the sensor monitoring system. 
After deriving the \ac{AoI} violation probability, 
we also present the $\eta$-coverage probability, which is the probability that the ETS coverage is greater than $\eta$ ratio of the maximum sensor coverage. 
Finally, we formulate the optimization problem for minimizing the average energy consumption of the sensor  
while guaranteeing certain level of $\eta$-coverage probability, and provide the optimal transmission power. }
{The main contributions of this paper can be summarized as follows:
\bi
\item we newly define the \ac{ETS} coverage, which reflects \emph{the spatial-temporal correlation} 
between the current exact information and the available information at the \ac{DC}, 
which is a function of the \ac{AoI} and the spatial distance from the sensing point; 
%
%
\item after deriving the \ac{AoI} violation probability, we present a \emph{closed-form expression of the $\eta$-coverage probability} and 
analytically show that the $\eta$-coverage probability increases with the transmission power of the sensor;
\item from the trade-off between the transmission power and the average number of retransmissions, 
we formulate the average energy consumption minimization problem, 
and provide \emph{the optimal transmission power of the sensor} for the noise-limited environments and the interference-limited environments; and
\item we analytically show the optimal transmission power increases with the maximum number of retransmissions regardless of system parameters' values. 
%
\ei
}} 
The remainder of this paper is organized as follows. In Section II, we present a network model, concept of the AoI and the $\eta$-coverage probability. In Section III, we derive the closed-form expression of the average AoI and the AoI violation probability. In Section IV, we formulate the average energy consumption minimization problem constrained on the target $\eta$-coverage probability. In Section V, we evaluate the theoretically obtained average AoI and AoI violation probability with the simulation and find the optimal transmission power to minimize the average energy consumption. Finally, the conclusion is presented in Section VI.

%
%
\section{System Model}

In this section, we describe the network model and the \ac{AoI} of the sensor monitoring system. We also present the sensing coverage model.
\begin{figure}[t!]
	\begin{center}   
		{ 
			\includegraphics[width=1\columnwidth]{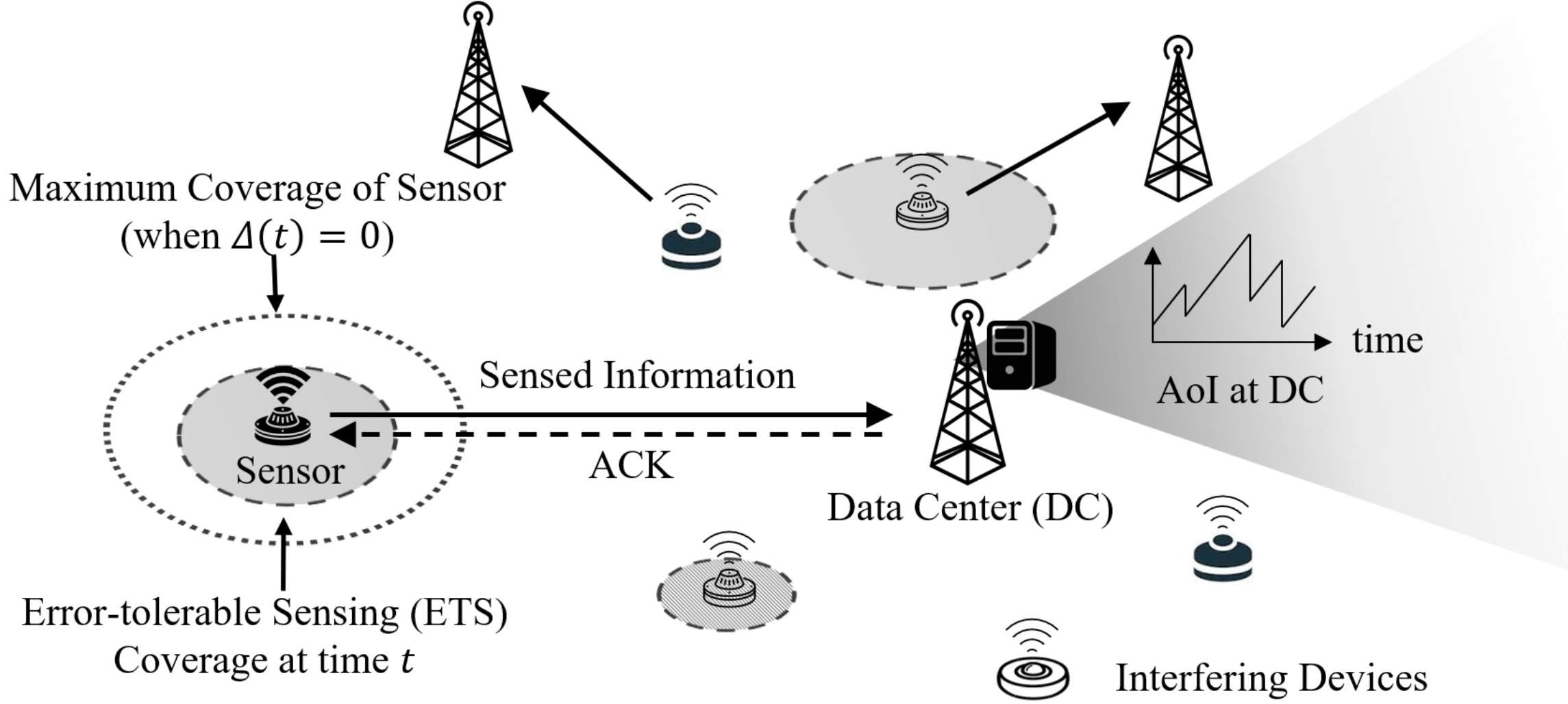}
			\vspace{-8mm}
		}
	\end{center}
	\caption{
		An example of the sensor monitoring system.
	}
	\vspace{-1mm}
	\label{fig:Sensor_model}
\end{figure}
\begin{figure}[t!]
	\begin{center}   
		{ 
			\includegraphics[width=0.90\columnwidth]{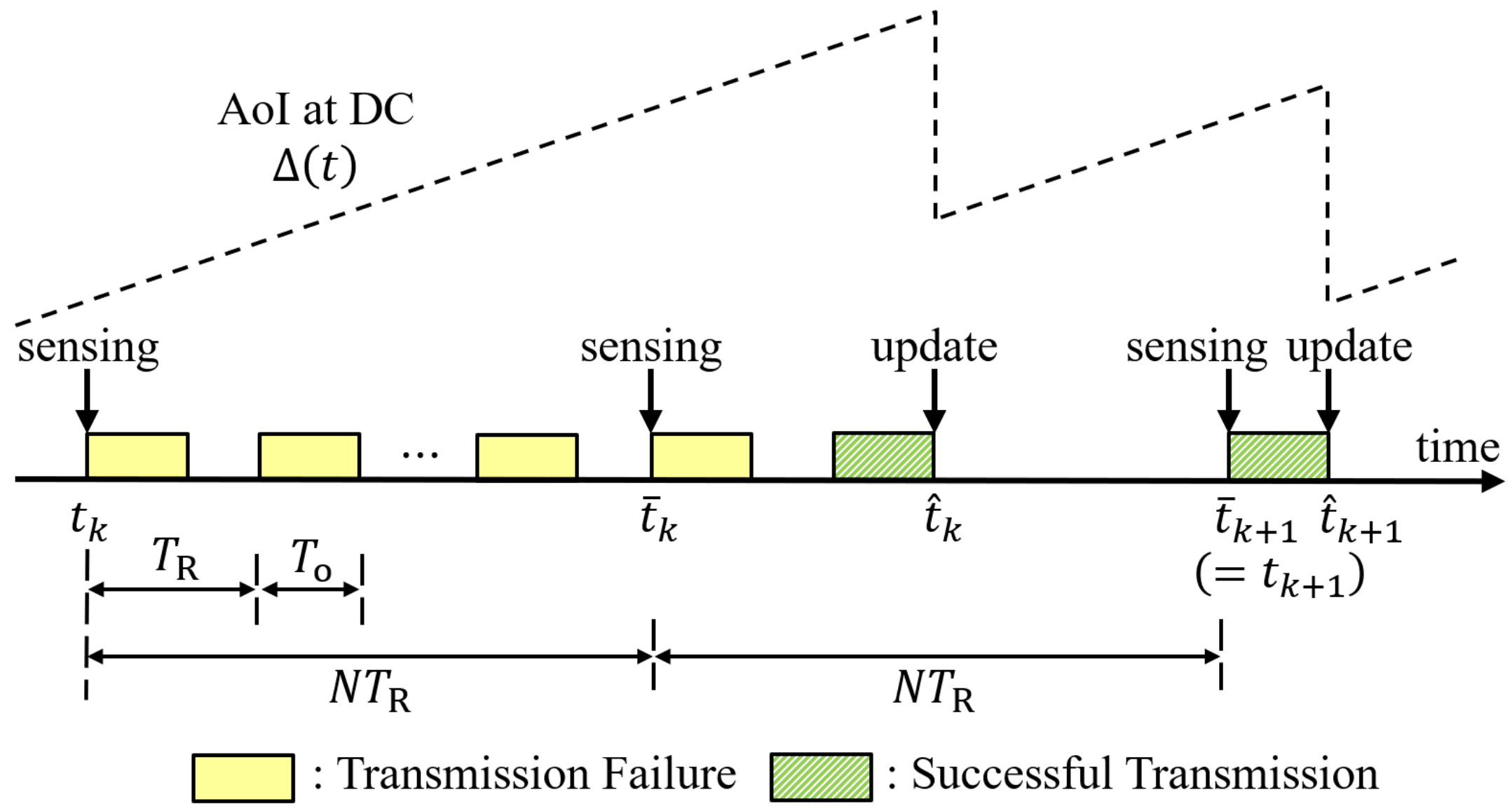}
			\vspace{-4mm}
		}
	\end{center}
	\caption{Procedure of the retransmission at a sensor and the change of AoI at the \ac{DC}.}
	\vspace{-1mm}
	\label{fig:AoI_procedure}
\end{figure}

\begin{table}
	\caption{Notations used throughout the paper.} \label{table:notation}
	\begin{center}
		\rowcolors{2}
		{cyan!15!}{}
		\renewcommand{\arraystretch}{1.4}
		\begin{tabular}{ c  p{7cm} }
			\hline 
			{\bf Notation} & {\hspace{2.5cm}}{\bf Definition}
			\\
			\midrule
			\hline
			$\SD$ 	& Successful transmission probability\\ 
			$\TransPower$ 	& Transmission power of a sensor \\ 
			$P_j$ 	& Transmission power of $j$-th type interfering IoT device \\ 
			$\CH$ 	& Channel fading gain \\ 
			$\distance$ 	& Distance of the link between a sensor and \ac{DC} \\ 
			$\PL$ 	& Pathloss exponent \\ 
			$\Noise$ 	& Noise power \\ 
			$\Interference$ 	& Interference at the \ac{DC} \\ 
			$\delta$ 	& Target SINR \\ 
			$\InterferenceDensity$ 	& Density of $j$-th type interfering IoT devices \\ 
			$\CommunicationTime$ 	& Transmission duration \\ 
			$\TransTime$ 	& Retransmission interval \\ 
			$\Ni$ 	& Maximum number of retransmission \\ 
			$\SuccTrans{k}$ 	& Time from the sensing instant after the $(k-1)$th successful update to the $k$th successful update \\ 
			$\RealSuccTrans{k}$ 	& Time of the $k$th successful update from its sensing instant \\ 
			$\WaitTime{k}$ 	& Waiting time for the next sensing after the $k$th successful update \\ 
			$\InterArea{k}$ 	& Trapezoid area between the $(k-1)$th and the $k$th successful updates in AoI\\
			$\IL{k}$ 	& Inter-arrival time between $(k-1)$th and $k$th successful updates \\ 
			$\Sensinginstant{k}$ 	& Sensing time after the $(k-1)$th successful update \\ 
			$\RealSensinginstant{k}$ 	& Sensing time of the $k$th successful update \\ 
			$\Arrivalinstant{k}$ 	& Time of the $k$th successful update \\ 
			$\Cov$ 	& Correlation of two information \\ 
			$\Error$ 	& Estimation error \\ 
			$\CoverageThres$ 	& Threshold of estimation error \\ 
			$\SensorCoverage$ 	& Maximum sensing coverage of a sensor \\ 
			$\Etacoverage$ 	& $\eta$-coverage probability \\ 
			$\AAoI$ 	& Average AoI \\ 
			$\PV{\Viol}$ 	& AoI violation probability \\ 
			$\VL{k}$ 	& Violated time of the $k$th update  \\ 
			$\Viol$ 	& AoI violation threshold \\ 
			$\EnergyTimeslot$ 	& Average energy consumption \\ 

			\hline 
		\end{tabular}
	\end{center}\vspace{-0.63cm}
\end{table}%
\subsection{Network Model}
 We consider a sensor monitoring system, which consists of sensors and the \ac{DC}, as shown in Fig. \ref{fig:Sensor_model}.
 Each sensor senses regional information (e.g., temperature, lighting, and humidity, etc.) periodically with the period $T_\text{s}$,
 and transmits it to the \ac{DC} via wireless communications during given transmission duration $\CommunicationTime$.  
The \ac{DC} maintains the information of the sensor, and updates it whenever it successfully receives the sensed information from the sensor. 
 
In this network, we assume there also exist IoT devices, which use the same frequency channel and interfere the communication of the sensor.
Specifically, we assume
 there are $J$-types of interfering IoT devices with indices $\mathcal{J}=\left\{1, 2, ..., J\right\}$,
 and the $j$-th type ($j \in \mathcal{J}$) interfering IoT devices are randomly distributed as a \ac{HPPP} $\Phi_j$ with the spatial density $\lambda_j$. 
 When the transmission power of the sensor is $\TransPower$, the SINR received by the \ac{DC} $y_0$ from the sensor $x_0$ can be expressed as
 \begin{align}\label{eq:SINR}
 	\text{SINR}_{x_0,y_0}
 	=
 	\frac{\TransPower h_{x_0,y_0}d_{x_0,y_0}^{-\PL}}{I+\Noise},
 \end{align}
where $\CH_{x,y}$ and $d_{x,y}$ are the channel fading gain and the distance of the link between nodes $x$ and $y$,
$\PL$ is the pathloss exponent, and 
$\Noise$ is the the power of the additive white Gaussian noise (AWGN). 
Here, $\Interference=\sum_{j\in\mathcal{J}}P_j\sum_{x\in\Phi_j}h_{x,y_0}d_{x,y_0}^{-\PL}$ is the received interference power,
where $P_j$ is the transmission power of the $j$-type interfering IoT devices.
We assume the channel experiences Rayleigh fading, i.e., $\CH_{x,y}\sim\exp(1)$ $\forall x,y$.


The transmission of a sensor can be successfully received at the \ac{DC}, 
when the SINR at the \ac{DC} is greater than or equal to the target SINR, $\delta$.
Then, the \ac{STP} can be presented as \cite{JeAn13}\footnote{This can be obtained by using 
	the Laplace transform of interference and the \ac{CDF} of the exponential distribution.}
%
\begin{align}\label{eq:STP}
\nonumber
\SD
&=
\Prob{\text{SINR}_{x_0,y_0}\geq\delta}
\\
&=
	\exp\left(-\frac{\xi}{\TransPower}\hspace{-0.5mm}-\frac{\zeta}{{\TransPower}^{2/\PL}}\hspace{-0.5mm}\right)
=
	\STPfunction,
\end{align}
%
%
%
where $\zeta$ and $\xi$ are given by 
\begin{align}	
	\zeta &={\delta}^{2/\PL}{\distance_{x_0,y_0}}^{2}\InterferenceGamma\sum_{j\in\mathcal{J}}\InterferenceDensity{P_j}^{2/\PL} \nonumber \\
	\xi & =\Noise\delta\distance_{x_0,y_0}^\PL \nonumber
\end{align}
for $\InterferenceGamma=\left( \frac{2\pi}{\PL}\right)\Gamma\left( \frac{2}{\PL}\right)\Gamma\left(1-\frac{2}{\PL}\right)$ 
with Gamma function $\Gamma\left(x\right)=\int_{0}^{\infty}t^{x-1}e^{-t}dt$.
From \eqref{eq:STP}, we can readily know that 
$\SD$ can be presented as a function of $\TransPower$, i.e., $\STPfunction$, which increases with $\TransPower$. 

Since the transmission from the sensor can be failed (with the probability $1-\SD$), 
 we consider the retransmission system, which allows the sensor to transmit up to $N$ times, as shown in Fig~\ref{fig:AoI_procedure}. 
 Specifically, the \ac{DC} sends an ACK signal to the sensor once it successfully receives the information. 
If the sensor does not receive the ACK signal for certain amount of time, it retransmits until it receives the ACK signal or it senses new information. 
Hence, when the (re)transmission can be happened in every $\TransTime$ (for $\TransTime > \CommunicationTime$), the sensing period is the same as $T_\text{s}= \SamplePeriod$.
%
 

%

\subsection{Age of Information}
The \ac{DC} updates the information of the sensor, 
when it receives the newly sensed information successfully. 
To measure the freshness of the sensor information at the \ac{DC} at time $t$, 
we use the \ac{AoI}, which is defined as\cite{KauYat12} 
%
\begin{align}
\AoI=t - t_\text{o} 
\end{align}
where $t_\text{o}$ is the sensing time of the most recently updated information at the \ac{DC}. 
As shown in Fig. \ref{fig:System_model}, we can see that the AoI increases linearly from the sensing instant over time, and then drops to the age of the updated information.

\subsection{$\eta$-Coverage Probability}

In the sensor monitoring networks, the sensed information is generally correlated in time and space \cite{Hricos19}. 
For instance, the sensed information of a sensor at time $t $ can have the correlation with that of the sensor at different location at a distance $l$ and time $t + \TD$.
We denote the correlation of those information as $\Correlation$ 
for the distance difference $l$ and the time difference $ \TD$ between sensors, 
which can be presented using the covariance model as \cite{Hricos19}
\begin{align}\label{eq:Covariance}
\Correlation=\exp\left\{-ul-v \TD\right\}
\end{align}
for the scaling parameters $u$ and $v$.
%
When those information have correlation as \eqref{eq:Covariance},
we can estimate the sensed information of a sensor at time $t + \TD$ 
from that of the other sensor at time $t $, and the estimation error can be presented as \cite{Hricos19}
\begin{align}\label{eq:EstimationError}
\epsilon\left(l, \TD\right)=1-\Correlation^2
=
1-\exp\left\{-2ul-2v\TD\right\}.
\end{align}


We now define the error-tolerable sensing (ETS) coverage of a sensor. 
The ETS coverage is the area that the sensor can provide valid information
with smaller error than $\CoverageThres$ at $t + \TD$ 
from its sensing information at $t$. 
From \eqref{eq:EstimationError}, the ETS coverage can be presented as a circle with the radius $\Radius$, given by
\begin{align}\label{eq:Radius2}
\nonumber
&\epsilon\left(l, \TD\right)\leq\CoverageThres
\\
&\Leftrightarrow \Radius\leq\frac{-\ln\left(1-\CoverageThres\right)-2v\TD}{2u}.
\end{align}
From \eqref{eq:Radius2}, we can see that $\Radius$ is the maximum 
when $\TD=0$ (i.e., at the sensing time), which is equal to 
$r_{\text{max}}\left(\CoverageThres\right)=\frac{-\ln\left(1-\CoverageThres\right)}{2u}$.
%
Then, $\Radius$ gets smaller over the time difference $\TD$, as shown in Fig. \ref{fig:Sensor_model}. 
This means when the AoI of the sensor's information at the \ac{DC} is $\AoI$ at time $t$,
the radius of the ETS coverage can be presented as
\begin{align}\label{eq:Radius}
r\left(\AoI, \CoverageThres\right)
=\sqrt{\frac{\SensorCoverage}{\pi}}-\frac{v}{u}\AoI
\end{align}
where $\SensorCoverage=\pi r^2_{\text{max}}\left(\CoverageThres\right)$ is the maximum ETS coverage.

Using \eqref{eq:Radius}, 
we now define the $\eta$-coverage probability $\Etacoverage$ 
as the probability that the ETS coverage is more than $\eta$ portion of the maximum ETS coverage $\SensorCoverage$, given by
\begin{align}\label{eq:Etacoverage}
\Etacoverage
&=
	\Prob{\frac{\pi\RadSquare}{\SensorCoverage}
		\geq
		\eta
	}.
\end{align}
Using \eqref{eq:Radius} and \eqref{eq:Etacoverage}, we can represent $\Etacoverage$ as
\begin{align}\label{eq:ETScoverage}
	\nonumber
	\Etacoverage
	&=
	\Prob{\AoI\leq\frac{u}{v}\sqrt{\frac{\SensorCoverage}{\pi}}\left(1-\sqrt{\eta}\right)}
	\\
	&=1-\PV{\frac{u}{v}\sqrt{\frac{\SensorCoverage}{\pi}}\left(1-\sqrt{\eta}\right)}
\end{align}
where $\PV{\Viol}=\Prob{\AoI>\Viol}$ is the AoI violation probability, which will be derived in the following section. This metric can show how large sensing area of the regional information the sensor has with smaller error than $\CoverageThres$.

\section{Age of Information Analysis}
In this section, we analyze the average AoI and the AoI violation probability of the sensor monitoring system.
%
%
%
\begin{figure}[t!]
	\begin{center}   
		{ 
			\includegraphics[width=0.80\columnwidth]{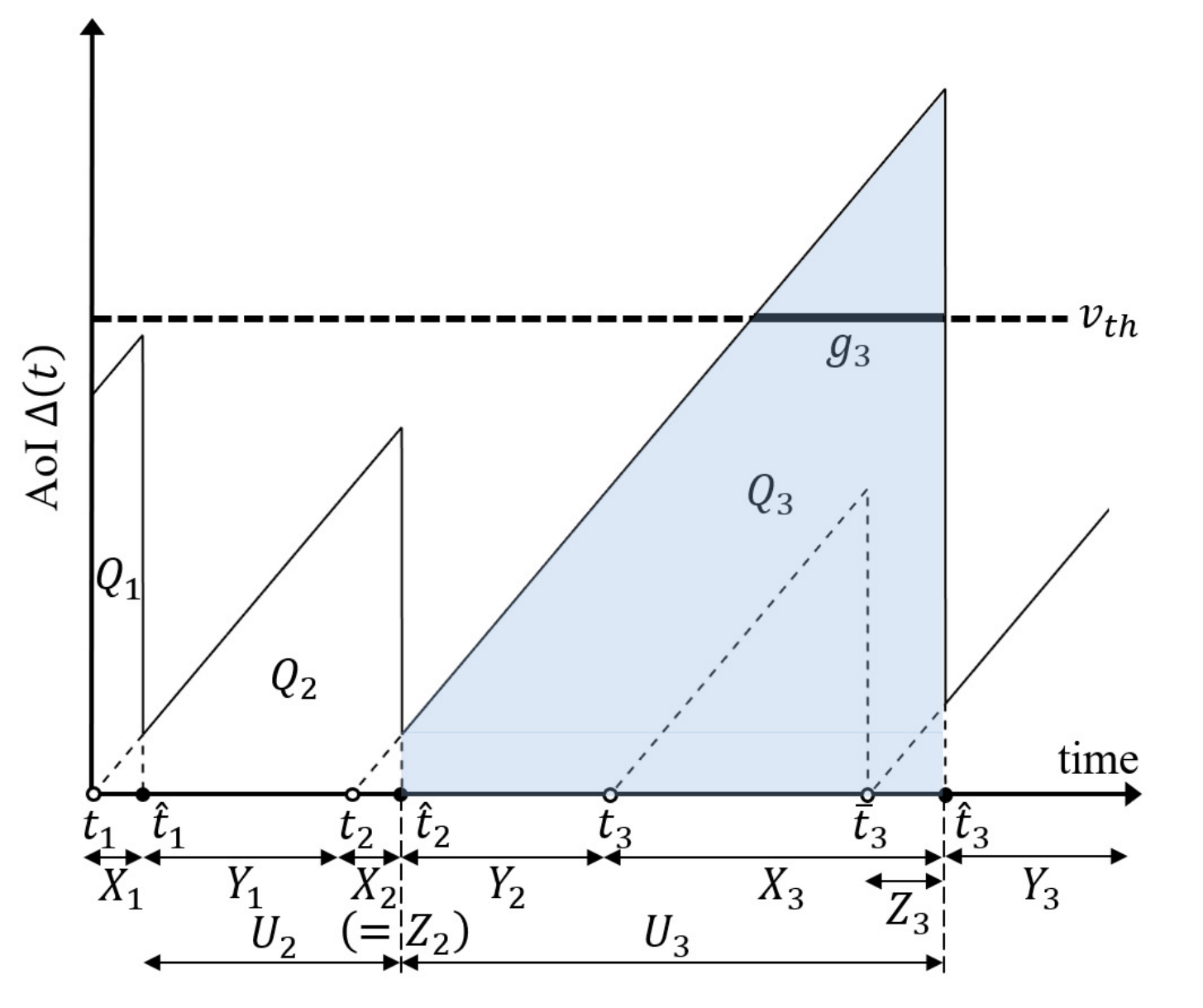}
			\vspace{-4mm}
		}
	\end{center}
	\caption{An example of AoI path at a \ac{DC}}
	\vspace{-1mm}
	\label{fig:System_model}
\end{figure}
%
%
For the analysis, we first define some notations, which is also used to present the AoI path at the \ac{DC} in Fig.~\ref{fig:System_model}.

We denote $\Sensinginstant{k}$ as the first sensing time after the $\left(k-1\right)$th successful update, 
$\RealSensinginstant{k}$ as the sensing time of the $k$th successful update, and $\Arrivalinstant{k}$ as the time of the $k$th successful update at the \ac{DC}.
Note that $\Sensinginstant{k}$ and $\RealSensinginstant{k}$ can be either the same or different (e.g., $\Sensinginstant{2}=\RealSensinginstant{2}$ in Fig.~\ref{fig:System_model}).
After the $\left(k-1\right)$th successful update,
the interval from the first sensing time $\Sensinginstant{k}$ to the $k$th successful update time $\Arrivalinstant{k}$ is denoted 
as  $\SuccTrans{k}=\Arrivalinstant{k}-\Sensinginstant{k}$.
%
Since $\SuccTrans{k}$ is determined by the number of retransmissions until the successful reception, 
its distribution can be modeled as the geometry distribution, of which the \ac{PMF} of $\SuccTrans{k}$ is given by
\begin{align}\label{eq:SuccessfulTransmissionProbability}
\Prob{\SuccTrans{k}\hspace{-0.7mm}
	=
		\hspace{-0.7mm}
		\left(n\hspace{-0.7mm}-\hspace{-0.7mm}1\right)\TransTime+\CommunicationTime}
	=
		\left(1-\SD\right)^{n-1}\SD,\ n=1,2, ... \ .
\end{align}
%
%
For the $k$th successful update, we denote 
the interval from its sensing time $\RealSensinginstant{k}$ to the successful update time $\Arrivalinstant{k}$
as $\RealSuccTrans{k}=\Arrivalinstant{k}-\RealSensinginstant{k}$. 
The distribution of $\RealSuccTrans{k}$ also follows the geometric distribution, but it is truncated by the maximum number of the retransmission $\Ni$ as\footnote{As $\RealSuccTrans{k}$ is defined from the sensing time of the specific $k$th successful update, it cannot be greater than 
$(\Ni-1)\TransTime+\CommunicationTime$.}
%
%
%
\begin{align}
\Prob{\RealSuccTrans{k}\hspace{-0.7mm}=\hspace{-0.7mm}\left(n\hspace{-0.7mm}-\hspace{-0.7mm}1\right)\TransTime\hspace{-0.7mm}+\hspace{-0.7mm}\CommunicationTime}
=
\frac{\left(1\hspace{-0.5mm}-\hspace{-0.5mm}\SD\right)^{n-1}\hspace{-0.5mm}\SD}{1\hspace{-0.5mm}-\hspace{-0.5mm}\left(1\hspace{-0.5mm}-\hspace{-0.5mm}\SD\right)^\Ni},\ n=1, ..., \Ni. \label{eq:prob_z}
\end{align}
After the $k$th successful update at $\Arrivalinstant{k}$, 
the time to the next sensing, i.e., $\Sensinginstant{k+1}$, is denoted as $\WaitTime{k}$, given by 
\begin{align}	
	\WaitTime{k}=\Sensinginstant{k+1}-\Arrivalinstant{k}=\SamplePeriod-\RealSuccTrans{k}. 
\end{align}
%
%
We then denote the interval between the $\left(k-1\right)$th and the $k$th successful updates at the \ac{DC} as $\IL{k}$, given by 
\begin{align}	
	\IL{k}
	=
		\Arrivalinstant{k+1} - \Arrivalinstant{k}
	=
		\WaitTime{k-1}+\SuccTrans{k}.
\end{align}
%
%
\subsection{Average AoI}
We first analyze the average AoI of the system. The average AoI at the \ac{DC} can be given by \cite{Mollei19}
\begin{align}\label{eq:AverageAoI}
\AAoI
&=
\lim\limits_{T\rightarrow\infty}\frac{1}{T} \int_{0}^{T}\AoI dt
\nonumber
\\
\nonumber
&=
\lim\limits_{T\rightarrow\infty}\left\{\frac{\InterArea{1}+\InterArea{\text{R}}}{T}+\frac{\numberofupdates-1}{T}\frac{1}{\numberofupdates-1}\sum_{k=2}^{\numberofupdates}\InterArea{k}\right\}
\\
&\approx
\frac{\Expt{\InterArea{k}}}{\Expt{\IL{k}}},
\end{align}
where $\InterArea{k}$ is the trapezoid area between the $\left(k-1\right)$th and the $k$th successful updates, $\numberofupdates=\max\left\{k\mid\Arrivalinstant{k}\leq T\right\}$ is the number of the successful updates until time $T$ and $\InterArea{\text{R}}=\frac{1}{2}\left(2\RealSuccTrans{\numberofupdates}+T-\Arrivalinstant{\numberofupdates}\right)\left(T-\Arrivalinstant{\numberofupdates}\right)$ is the residual area after the last update to $T$.
In the following Lemma, using \eqref{eq:AverageAoI}, we obtain the average \ac{AoI}.
\begin{lemma} \label{thm:AoI_avg}
	The average AoI of the sensor monitoring system can be expressed as
	\begin{align}
	\AAoI
	=
	\TransTime\left\{\frac{1}{\SD}+\frac{\Ni-2}{2}\right\}+\CommunicationTime.
	\label{eq:AAoI}
	\end{align}
\end{lemma}
\begin{figure*}[t!]
	\vspace*{0pt}
	\normalsize
	\begin{align}
	\label{eq:SecondViolationProbability}
	& \PV{\Viol}=
	\\
	\nonumber
	&\left\{
	\begin{aligned}
	&\frac{\left\{\SD\left(\CommunicationTime\hspace{-0.5mm}-\hspace{-0.5mm}\TransTime\hspace{-0.5mm}-\hspace{-0.5mm}\Viol\right)\hspace{-0.5mm}+\hspace{-0.5mm}\TransTime\right\}\hspace{-0.5mm}\left\{1\hspace{-0.5mm}-\hspace{-0.5mm}{\CP}^{\NV}\right\}\hspace{-0.5mm}+\hspace{-0.5mm}\SD\TransTime\hspace{-0.5mm}\left\{\Ni\hspace{-0.5mm}-\hspace{-0.5mm}\left(\NV\right)\hspace{-0.7mm}{\CP}^{\NV}\right\}}{\SD\SamplePeriod}, & \text{if } \Viol\leq\SamplePeriod,
	\\
	&\frac{\left\{\SD\TransTime\hspace{-0.5mm}\left\lfloor\frac{\Viol-\CommunicationTime}{\TransTime}\right\rfloor+\TransTime+\SD\CommunicationTime-\SD\Viol\right\}\left\{\CP^{\left\lfloor\frac{\Viol-\CommunicationTime}{\TransTime}\right\rfloor-\Ni+1}
		-
		\left(1-\SD\right)^{\left\lfloor\frac{\Viol-\CommunicationTime}{\TransTime}\right\rfloor+1}
		\right\}}{\SD\SamplePeriod}
	, & \text{if } \Viol>\SamplePeriod.
	\end{aligned}
	\right.
	\end{align}
	\centering \rule[4pt]{18cm}{0.3pt}
\end{figure*}
%
%
%
\begin{IEEEproof}
	From Fig. \ref{fig:System_model}, the expectations of $\InterArea{k}$ and $\IL{k}$ in \eqref{eq:AverageAoI} are, respectively, expressed as
	%
	%
	\begin{align}\label{eq:ExpectationArea}
	\Expt{\InterArea{k}}
	&=
	\Expt{\frac{1}{2}\left(\WaitTime{k-1}+\SuccTrans{k}\right)\left(2\RealSuccTrans{k-1}+\WaitTime{k-1}+\SuccTrans{k}\right)}
	\nonumber \\
	&=
	\Expt{\frac{1}{2}\left(\SamplePeriod+\SuccTrans{k}-\RealSuccTrans{k-1}\right)\left(\SamplePeriod+\SuccTrans{k}+\RealSuccTrans{k-1}\right)}
	\nonumber \\
	&\mathop=\limits^{\mathrm{(a)}}
	\frac{1}{2}\Expt{\left(\SamplePeriod+\SuccTrans{k}\right)^2}-\frac{1}{2}\Expt{\RealSuccTrans{k}^2},
	\\
	\label{eq:ExpectationInterval}
	\Expt{\IL{k}}
		&=
	\Expt{\WaitTime{k-1}+\SuccTrans{k}}
	=\Expt{\SamplePeriod-\RealSuccTrans{k-1}+\SuccTrans{k}}
	\nonumber \\
	&\mathop=\limits^{\mathrm{(b)}}
	\SamplePeriod-\Expt{\RealSuccTrans{k}}+\Expt{\SuccTrans{k}}
	\end{align}
	where (a) and (b) follows since $\RealSuccTrans{k-1}$ and $\RealSuccTrans{k}$ are identically distributed. 
	From \eqref{eq:SuccessfulTransmissionProbability}, $\Expt{\SuccTrans{k}}$ and $\Expt{\SuccTrans{k}^2}$ are obtained, respectively, as
	\begin{align}\label{eq:ExpectationX}
	\Expt{\SuccTrans{k}}
	&=
	\sum_{n=1}^{\infty}\left\{\left(n-1\right)\TransTime+\CommunicationTime\right\} \left(1-\SD\right)^{n-1}\SD
	\nonumber \\
	&=
	\left(\frac{1}{\SD}-1\right)\TransTime+\CommunicationTime,
	\\
	\label{eq:ExpectationXsquare}
	\nonumber
	\Expt{\SuccTrans{k}^{2}}
	&=
	\sum_{n=1}^{\infty}\left\{\left(n-1\right)\TransTime+\CommunicationTime\right\}^2 \left(1-\SD\right)^{n-1}\SD
	\\
	&=\frac{2\TransTime^2}{\SD^2}-\frac{\left(3\TransTime-2\CommunicationTime\right)\TransTime}{\SD}+\left(\TransTime-\CommunicationTime\right)^2.
	\end{align}
	Similarly, from \eqref{eq:prob_z}, we obtain $\Expt{\RealSuccTrans{k}}$ and $\Expt{\RealSuccTrans{k}^{2}}$, respectively, as
	\begin{align}\label{eq:ExpectationZ}
	\Expt{\RealSuccTrans{k}}
	&=
	\sum_{n=1}^{N}\left\{\left(n-1\right)\TransTime+\CommunicationTime\right\}\frac{\left(1-\SD\right)^{n-1}\SD}{1-\left(1-\SD\right)^N}
	\nonumber \\
	& =\frac{\TransTime\left\{1-\CP^{\Ni}\left(1+\SD\Ni\right)\right\}}{\SD\left\{1-\CP^{\Ni}\right\}}-\TransTime+\CommunicationTime,
	\end{align}
	\begin{align}\label{eq:ExpectationZsquare}
		&\Expt{\RealSuccTrans{k}^2}
	=
	\sum_{n=1}^{N} \left\{\left(n - 1\right) \TransTime + \CommunicationTime\right\}^2 \frac{\left(1 - \SD\right)^{n-1} \SD} {1 - \left(1 - \SD\right)^N}
	\nonumber \\
	&\mathop = \limits^{\mathrm{(a)}}
	\left(\TransTime\hspace{-0.5mm}-\hspace{-0.5mm}\CommunicationTime\right)^2\hspace{-0.4mm}+
	\frac{\TransTime}{\SD\left\{1-\left(1-\SD\right)^N\right\}}\hspace{-0.5mm}\left[\frac{{\TransTime}}{\SD}\hspace{-0.5mm}\left\{-N^2\hspace{-0.5mm}\left(1-\SD\right)^{N+2}
	\right. \right.	\nonumber \\
	&+\left(2N^2+2N-1\right)\left(1-\SD\right)^{N+1}
	\left. 
	- \left(N\hspace{-0.5mm}+\hspace{-0.5mm}1\right)^2\left(1\hspace{-0.5mm}-\hspace{-0.5mm}\SD\right)^N-\SD\right.
	\nonumber \\
	&\left.+2\right\}\left.-2\hspace{-0.5mm}\left(\TransTime\hspace{-0.7mm}-\hspace{-0.7mm} \CommunicationTime\right)\hspace{-0.5mm}\left\{\hspace{-0.5mm} 1\hspace{-0.7mm}-\hspace{-0.7mm}\left(1\hspace{-0.7mm}-\hspace{-0.7mm}\SD\right)^{\Ni} \hspace{-1.2mm} \left(1\hspace{-0.7mm}+\hspace{-0.5mm}\SD\Ni\right)\hspace{-0.5mm}\right\} \hspace{-0.5mm}\right]
	\end{align}
	where (a) is obtained from the equation in \cite[eq. (0.114)]{GraRyz07}.
	By substituting \eqref{eq:ExpectationX}, \eqref{eq:ExpectationXsquare} and \eqref{eq:ExpectationZsquare} into \eqref{eq:ExpectationArea}, and \eqref{eq:ExpectationX} and \eqref{eq:ExpectationZ} into \eqref{eq:ExpectationInterval}, $\Expt{\InterArea{k}}$ and $\Expt{\IL{k}}$ are presented, respectively, as
	\begin{align}
	\Expt{\InterArea{k}}\label{eq:ExpectedInterArea}
	=&
	\frac{\Ni\TransTime\left[\left\{2+\left(\Ni-2\right)\SD\right\}\TransTime+2\SD\CommunicationTime\right]}{2\SD\left\{1-\left(1-\SD\right)^{\Ni}\right\}},
	\\
	\Expt{\IL{k}}\label{eq:Interarrival}
	=&
	\frac{\SamplePeriod}{\CPP}.
	\end{align}
	Finally, substituting \eqref{eq:ExpectedInterArea} and \eqref{eq:Interarrival} into \eqref{eq:AverageAoI} results in \eqref{eq:AAoI}.
\end{IEEEproof}
From Theorem~\ref{thm:AoI_avg}, we can see that the average AoI increases linearly with the retransmission duration $\TransTime$ and the maximum number of the retransmissions $\Ni$. 
In addition, we can also see that the average AoI decreases with $\SD$, since $\frac{d}{d\SD}\AAoI=-\frac{\TransTime}{\SD^2}$ is negative as $\TransTime$ and $\SD$ are positive.
\subsection{AoI Violation Probability}
In this subsection, we derive the closed-form expression of the AoI violation probability. The AoI violation probability can be defined as \cite{Chazub19}
\begin{align}
\PV{\Viol}\label{eq:ViolatedLength}
=\Prob{\AoI>\Viol}
=\lim_{T\rightarrow\infty}\frac{1}{T}\sum_{k=1}^{L\left(T\right)}\VL{k}
=
\frac{\Expt{\VL{k}}}{\Expt{\IL{k}}},
\end{align}
where $\VL{k}$ is the violated time duration of the $k$th successful update (as illustrated in Fig. \ref{fig:System_model}).
\begin{theorem} \label{thr:APP}
	The AoI violation probability $\PV{v_\text{th}}$ of sensor monitoring system is given in \eqref{eq:SecondViolationProbability}, as shown at the top of this page, where $\lfloor\cdot\rfloor$ is a floor function. \label{thm:violation}
\end{theorem}
		\begin{IEEEproof}
			See Appendix \ref{App:thm1}.
		\end{IEEEproof}
	From Theorem~\ref{thr:APP}, we can express the $\eta$-coverage probability in \eqref{eq:ETScoverage} as the closed-form. In addition, we obtain the following corollary.
\begin{corollary} \label{col:inequality}
The AoI violation probability decreases with the STP $\SD$, i.e., $\frac{d\PV{\Viol}}{d\SD}\leq 0$.
\end{corollary}
\begin{IEEEproof}
	Firstly, when $\Viol\leq\SamplePeriod$, the derivative of $\PV{\Viol}$ with respect to $\SD$ is given by the first equation in \eqref{eq:SecondDerivativeViolation},
\begin{figure*}[t!]
	\vspace*{4pt}
	\normalsize
	\begin{align}
	\label{eq:SecondDerivativeViolation}
	&\frac{d\PV{\Viol}}{d\SD}=
	\\
	&\left\{
	\begin{aligned}
	&\frac{\CP^{\left\lfloor\hspace{-0.5mm}{\frac{\Viol-\CommunicationTime}{\TransTime}}\hspace{-0.5mm}\right\rfloor}\left\{\TransTime\left(1-\SD\right)+\left(\NV\right)\SD\brep-\left(\NV\right)\SD^{2}\left(\TransTime-\CommunicationTime\right)\right\}-\TransTime}{\SamplePeriod{\SD}^2}
	, \qquad\qquad\qquad\text{if } \Viol\leq\SamplePeriod,
	\\
	\nonumber
	&\frac{\overbrace{\CP\left[\CP^{\Ni-1}\left\{\TransTime\CP+\Ni\SD\drep-\Ni\SD^2\left(\TransTime-\CommunicationTime\right)\right\}-\TransTime\right]}^{\text{(a)}}\overbrace{-\frep\SD\left(\drep+\SD\CommunicationTime-\SD\TransTime\right)\left\{\CPP\right\}}^\text{(b)}}{\SamplePeriod \SD^{2} {\CP}^{1-\frep}},  \\&\qquad\qquad\qquad\qquad\qquad\qquad\qquad\qquad\qquad\quad\qquad\qquad\qquad\qquad\qquad\qquad\qquad\qquad\qquad\qquad\qquad\qquad \text{if } \Viol\hspace{-0.5mm}>\hspace{-0.5mm}\SamplePeriod.
	\end{aligned}
	\right.
	\end{align}
	\centering \rule[0pt]{18cm}{0.3pt}
\end{figure*}
	where $\brep=\TransTime-\SD\arep$ and $\arep=\Viol-\left(\NV\right)\TransTime$ with $0\leq\arep\leq\TransTime$, then $\CP\TransTime\leq\brep\leq\TransTime$. 
%
%
%
%
	Here, we define the numerator of the first equation in \eqref{eq:SecondDerivativeViolation} as $\FirstNominator{\SD, \beta_1}$, which has the maximum value when $\brep=\TransTime$.
	This is because $\FirstNominator{\SD, \beta_1}$ increases with $\brep$.
	We then obtain the derivative of $\FirstNominator{\SD, \TransTime}$ with respect to $\SD$ as
	\begin{align}\label{eq:DerivativeNominator}
	&\frac{d}{d\SD}\FirstNominator{\SD, \TransTime}
	\hspace{-0.7mm} = \hspace{-0.7mm}
	-\left(\NV\right) \hspace{-0.7mm} \SD{\left(1 \hspace{-0.7mm} - \hspace{-0.7mm} \SD\right)}^{\left\lfloor\hspace{-0.5mm}{\frac{\Viol-\CommunicationTime}{\TransTime}}\hspace{-0.5mm}\right\rfloor-1} 
	\nonumber \\
	&\left\{ \hspace{-0.7mm}\left\lfloor\hspace{-0.5mm}{\frac{\Viol\hspace{-0.5mm}-\hspace{-0.5mm}\CommunicationTime}{\TransTime}}\hspace{-0.5mm}\right\rfloor \hspace{-0.3mm} \CommunicationTime
	+ \left(\TransTime\hspace{-0.5mm}-\hspace{-0.5mm}\CommunicationTime\right)\left(\left\lfloor\hspace{-0.5mm}{\frac{\Viol\hspace{-0.5mm}-\hspace{-0.5mm}\CommunicationTime}{\TransTime}}\hspace{-0.5mm}\right\rfloor+2\right)\left(1-\SD\right)\right\}
	\leq 0,	
	\end{align}
	where $\TransTime-\CommunicationTime\geq0$.
	From \eqref{eq:DerivativeNominator} and the fact that $\FirstNominator{\SD, \beta_1}$ increases with $\brep$, $\FirstNominator{\SD, \brep}$ 
	is not higher than $\FirstNominator{\SD, \TransTime}$. 
	Moreover, since $\FirstNominator{0, \brep}=0$, we can see that $\FirstNominator{\SD, \brep}$ is negative. Therefore, $\PV{\Viol}$ monotonically decreases with $\SD$. 
	On the other hand, when $\Viol>\SamplePeriod$, the derivative of $\PV{\Viol}$ with respect to $\SD$ is given by \eqref{eq:SecondDerivativeViolation},
	where $\drep=\TransTime\left(1+\Ni\SD\right)-\SD\left(\Viol-\frep\TransTime\right)$ with $\gamma=\NVN+1\geq 1$, then $\TransTime\leq\drep<\TransTime\left(1+p\right)$. 
	The numerator of the second equation in \eqref{eq:SecondDerivativeViolation} 
	has two parts: (a) and (b).
	The fact that (a) is negative can be proven in the same way as the first case (i.e., $\Viol\leq\SamplePeriod$). 
	%
	%
	%
	%
	%
	%
	In addition, 
	(b) is negative due to $\drep+\SD\CommunicationTime-\SD\TransTime>0$.
	Therefore, $\PV{\Viol}$ monotonically decreases with $\SD$. 
	%
\end{IEEEproof}

From Corollary~1, it can be seen that the AoI violation probability decreases as the transmission power increases or the distance between the sensor and \ac{DC} decreases.
\section{Average Energy Consumption Minimization}
 In this subsection, we consider the average energy consumption minimization problem of the sensor monitoring system.
 The average energy consumption of the system $\mathcal{E}\left(\TransPower\right)$ per retransmission interval $\TransTime$ is given by
 \begin{align}\label{eq:Energypertimeslot}
 \EnergyTimeslot
 =
 \frac{\Es+\bar{N}\TransPower\CommunicationTime}{\Ni},
 \end{align}
 where $\Es$ is the sensing energy and $\bar{N}$ is the average number of retransmissions per the sensing period $\SamplePeriod$, which can be represented as
 \begin{align}\label{eq:AverageTransmission}
 \bar{N}
 &=
 \sum_{n=1}^{\Ni}n\CP^{n-1}\SD+\Ni\CP^{\Ni}
 \nonumber \\
 &=
 \frac{1-\CP^{\Ni}}{\SD}.
 \end{align}
 Note that, in \eqref{eq:Energypertimeslot}, there is a trade-off between the transmission energy $P_\text{t}T_\text{o}$ and the average number of retransmissions $\bar{N}$. 
 Specifically, as the transmission power $P_\text{t}$ increases, $P_\text{t}T_\text{o}$ increases, but $\bar{N}$ decreases because $\frac{d \bar{N}} {d p_\text{s}} < 0$ in \eqref{eq:AverageTransmission}. 
 Therefore, we consider the problem of the average energy consumption minimization by optimizing $\TransPower$ while guaranteeing the $\eta$-coverage probability not less than the target probability $\epsilon$ as follows.
  	\begin{problem}[Average Energy Consumption Minimization Problem]\label{eq:MinEnergy}
  	\begin{alignat}{2}
  	\!\min_{\TransPower} \quad 
  	& 
  	\frac{\Es}{\Ni}+\frac{\CommunicationTime\TransPower\left\{\CPP\right\}}{\Ni\SD}
  	\nonumber
  	\\
  	%
  	\text{s.t.} \quad
  	&
	\nonumber
	\Etacoverage\geq\epsilon.
  	\end{alignat}
  \end{problem}
 Note that the objective function of Problem~1 is $\EnergyTimeslot$, which is presented using \eqref{eq:Energypertimeslot} and \eqref{eq:AverageTransmission}. From \eqref{eq:ETScoverage}, we can also present the constraint using the AoI violation probability as
  \begin{align}\label{eq:ModifiedViolationProb}
  &\mathcal{P}_{\text{c}}\left(\eta\right) \geq \epsilon 
  \nonumber \\
  &\Leftrightarrow \PV{\frac{u}{v}\sqrt{\frac{\SensorCoverage}{\pi}}\left(1-\sqrt{\eta}\right)}
  \leq
  1-\epsilon
  \end{align}
 where $\PV{\Viol}$ is given in \eqref{eq:SecondViolationProbability}. Furthermore, $\TransPower$ can also be presented as a function of $\SD$, i.e., $\TransPower=\STPinvfunction$, where $\STPfunction$ is defined in \eqref{eq:STP}. Since the function $\STPfunction$ is bijective, we can make the equivalent problem of Problem~1 as a minimization problem over $\SD$ instead of $\TransPower$. 

In addition, in Corollary~1, we show that $\PV{\Viol}$ is a decreasing function of $\SD$. If we define the AoI violation probability in \eqref{eq:SecondViolationProbability} as a function of $\SD$, i.e., $\PV{\SD}$ for better presentation, the constraint in \eqref{eq:ModifiedViolationProb} can be represented as\footnote{Unfortunately, it is hard to obtain the exact value of $\PVinv{1-\epsilon}$ in a closed-form expression, but it can be readily obtained numerically.}
\begin{align}\label{eq:MinimumCoverageSTP}
	\PV{\SD}\leq 1-\epsilon
	\Leftrightarrow
	\SD\geq\PVinv{1-\epsilon}.
\end{align}

Therefore, the average energy consumption minimization problem can be finally represented as
\begin{problem}[Equivalent problem to Problem 1]\label{eq:MinEnergy3}
	\begin{alignat}{2}
	\!\min_{\SD} \quad 
	& 
	\frac{\Es}{\Ni}+\frac{\CommunicationTime\STPinvfunction\left\{\CPP\right\}}{\Ni\SD}
	=
	\STPEnergyTimeslot
	\nonumber
	\\	  
	\text{s.t.} \quad
	&
	\SD\geq\PVinv{1-\epsilon}
	=
	\CoverageSol
	.
\nonumber
	\end{alignat}
\end{problem}
Note that $\STPinvfunction$ cannot be presented in a closed form for general environments.
Hence, we consider two environments for the optimization: the noise-limited and the interference-limited environments. 
From \eqref{eq:STP}, we can then present $\STPinvfunction$ in a closed form as
%
%
	\begin{numcases}{\STPinvfunction=}
	\label{eq:STPinvSNR}
	-\frac{\NoiseCoefficient}{\ln\SD},\,\,\,\,\,\,\,\,\,\,\,\,\,\,\,\,\,\,\,\,\,\,\,\,\, \text{noise-limited} 
	\\
	\label{eq:STPinvSIR}
	\InterferenceCoefficient\left(-\frac{1}{\ln\SD}\right)^{\PL/2},\,\, \text{interference-limited.}
	\end{numcases}
After presenting $\STPinvfunction$ as \eqref{eq:STPinvSNR} and \eqref{eq:STPinvSIR}, we can see that the objective function $\STPEnergyTimeslot$ is not a convex, 
so it is hard to obtain the optimal solution. 
However, since the objective function becomes differentiable, 
we first investigate the stationary points of the objective function, satisfying the constraint as follows.



\begin{figure}[t!]
	\begin{center}   
		{ 
			\psfrag{X}[bl][bl][0.7] {Path loss coefficient $\PL$}
			\psfrag{Y}[bl][bl][0.7] {$\FirstN$}
			\includegraphics[width=1.00\columnwidth]{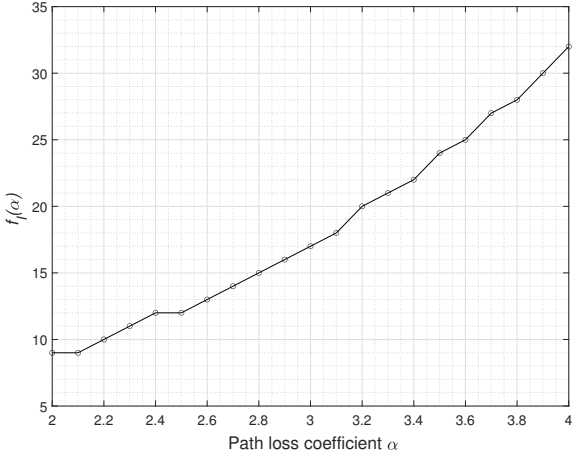}
		}
	\end{center}
	\caption{Minimum value of $\Ni$ which gives two stationary points $\FirstN$ as a function of path loss coefficient $\PL$.}
	\vspace{0.7mm}
	\label{fig:alpha_vs_N}
\end{figure}

\begin{proposition}
	Regardless of system parameter values, 
	$\STPEnergyTimeslot$ has two stationary points for the following range of the maximum retransmission number $\Ni$: 
	\begin{itemize}
		\item $\Ni\geq9$ in the noise-limited environment
		\item $\Ni\geq\FirstN$ in the interference-limited environment
	\end{itemize}
	where the values of $\FirstN$ are provided in Fig. \ref{fig:alpha_vs_N} for the pathloss exponent $2\leq \alpha \leq 4$. 
	In other range of $\Ni$, $\STPEnergyTimeslot$ does not have any stationary point. 
	
\end{proposition}
\begin{IEEEproof}
	In the noise-limited environment, by substituting \eqref{eq:STPinvSNR} into \eqref{eq:Energypertimeslot}, $\STPEnergyTimeslot$ is presented as
	\begin{align}\label{eq:EnergyEnergy}
	\STPEnergyTimeslot
	=
	\frac{\Es}{\Ni}-\frac{\NoiseCoefficient\CommunicationTime\left\{1-\CP^{\Ni}\right\}}{\Ni\SD\ln\SD}.
	\end{align}
	From \eqref{eq:EnergyEnergy}, we obtain the derivative of $\STPEnergyTimeslot$ with respect to $\SD$ as
	\begin{align}\label{eq:FirstEnergyDerivative}
	\frac{d\STPEnergyTimeslot}{d\SD}
	=
	\frac{\NoiseCoefficient \CommunicationTime}{\Ni}\EnergyDerivativeSNR
	\end{align}
	where
	\begin{align}
	\EnergyDerivativeSNR
	\hspace{-0.3mm}=\hspace{-0.3mm}
	\frac{\left(1\hspace{-0.5mm}+\hspace{-0.5mm}\ln\SD\right) \hspace{-0.8mm} \left\{1\hspace{-0.7mm}-\hspace{-0.7mm} \left(1\hspace{-0.7mm}-\hspace{-0.7mm}\SD\right)^{\Ni}\right\} \hspace{-0.7mm}-\hspace{-0.7mm}\Ni\SD{\left(1\hspace{-0.7mm}-\hspace{-0.7mm}\SD\right)}^{\Ni-1}\hspace{-0.5mm}\ln\SD} {{\SD}^2 \left(\ln\SD\right)^2}.
	\end{align}
	In \eqref{eq:FirstEnergyDerivative}, $\frac{\NoiseCoefficient \CommunicationTime}{\Ni}>0$ and $\EnergyDerivativeSNR$ is a function of the parameters $\SD$ and $\Ni$ only. Hence, the stationary points that make $\frac{d\STPEnergyTimeslot}{d\SD}=0$ are determined by $\EnergyDerivativeSNR=0$ and they are affected by $\Ni$ only.
	Nevertheless, for given $\Ni$, it is difficult to present the exact value of $p_s$ that makes $\EnergyDerivativeSNR=0$.
	However, by plotting $\EnergyDerivativeSNR$, we can readily find $\SD$ that makes $\EnergyDerivativeSNR=0$, which are the stationary points.
	In Fig. \ref{fig:Energy_derivative}, we plot $\EnergyDerivativeSNR$ for different values. 
	In this figure, we can see that if $\Ni\geq9$, there exist two stationary points, marked by circles, otherwise, there is no stationary point. 
	%
	%
	\begin{figure}[t!]
		\begin{center}   
			{ 
				\psfrag{AAAAAA}[bl][bl][0.7] {$\Ni=5$}
				\psfrag{A1}[bl][bl][0.7] {$\Ni=8$}
				\psfrag{A2}[bl][bl][0.7] {$\Ni=9$}
				\psfrag{A3}[bl][bl][0.7] {$\Ni=15$}
				\psfrag{X}[bl][bl][0.7] {$\SD$}
				\psfrag{Y}[bl][bl][0.7] {$\EnergyDerivativeSNR$}
				\psfrag{P1}[bl][bl][0.7] {$\FirstLocalOptimum$}
				\psfrag{P2}[bl][bl][0.7] {$\SecondLocalOptimum$}
				\includegraphics[width=1.00\columnwidth]{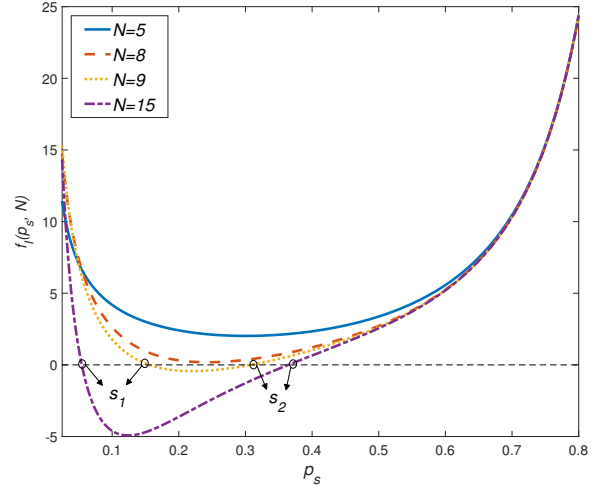}
				\vspace{-5mm}
			}
		\end{center}
		\caption{$\EnergyDerivativeSNR$ in the proof of Proposition~1 as a function of $p_s$ for different values of $\Ni$}
		\vspace{-3mm}
		\label{fig:Energy_derivative}
	\end{figure}
	
	In the interference-limited environment, by substituting \eqref{eq:STPinvSIR} into \eqref{eq:Energypertimeslot}, we can present $\STPEnergyTimeslot$ as
	\begin{align}\label{eq:EnergyEnergy2}
	\STPEnergyTimeslot
	=
	\frac{\Es}{\Ni}+\frac{\InterferenceCoefficient\CommunicationTime\left\{1-\CP^{\Ni}\right\}}{\Ni\SD\left(-\ln\SD\right)^{\PL/2}}.
	\end{align}
	In \eqref{eq:EnergyEnergy2}, we have
	\begin{align}\label{eq:SecondEnergyDerivative}
	&\frac{d\STPEnergyTimeslot}{d\SD}
	=
	\frac{\InterferenceCoefficient \CommunicationTime}{2\Ni}\EnergyDerivativeSIR
	\end{align}
	where $\EnergyDerivativeSIR$ is given by
	\begin{align}
	&\EnergyDerivativeSIR
	\nonumber
	\\
	&=
	\frac{\left(\hspace{-0.2mm} \PL \hspace{-0.7mm} + \hspace{-0.7mm} 2 \ln\SD \hspace{-0.3mm} \right) \hspace{-0.6mm} 
		\left\{ \hspace{-0.3mm} 1 \hspace{-0.7mm} - \hspace{-0.7mm} \left(\hspace{-0.2mm}1\hspace{-0.7mm}-\hspace{-0.7mm}\SD\hspace{-0.3mm}\right)^{\Ni} \hspace{-0.5mm} \right\} \hspace{-0.6mm}-\hspace{-0.5mm} 2\Ni\SD \hspace{-0.5mm} \left(\hspace{-0.2mm}1\hspace{-0.7mm}-\hspace{-0.7mm}\SD\hspace{-0.3mm}\right)^{\Ni-1} \hspace{-0.3mm} \ln\SD}
	{\SD^{2} \left(-\ln{\SD}\right)^{\frac{2+\PL}{2}}}.
	\end{align}
	In \eqref{eq:SecondEnergyDerivative}, $\EnergyDerivativeSIR$ is a function of $\SD$, $\Ni$ and $\PL$, so the stationary points are only affected by $\Ni$ and $\PL$. 
	Since $\frac{\InterferenceCoefficient T_\text{o}}{2\Ni}>0$, 
	$p_s$ which makes $\EnergyDerivativeSIR =0$ becomes the stationary point. 
	Similar to $\EnergyDerivativeSNR$, we can readily know that $\EnergyDerivativeSIR$ also gives two stationary points for larger $\Ni$ than a certain value $\FirstN$, which is differently determined by $\PL$. For smaller $\Ni < \FirstN$, there exits no stationary point. 
	After checking $\EnergyDerivativeSIR$ for different $\PL$, 
	we obtain the value $\FirstN$ for $2 \leq \PL \leq 4$ as Fig.~\ref{fig:alpha_vs_N}.
\end{IEEEproof}

From Proposition~1, 
	we can know that when $\Ni$ is small, $\STPEnergyTimeslot$ increases with $p_s$ since 
	$\frac{d\STPEnergyTimeslot}{d\SD}>0$. This means it cannot be energy-efficient if we use larger transmission power to increase the \ac{STP} in this case. We can also obtain the optimal value of the \ac{STP} in the following Lemma.

	%
	%
	%
\begin{lemma}  \label{lem:optimal}
	For $\Ni\geq9$ in the noise-limited environment and $\Ni\geq\FirstN$ in the interference-limited environment, the optimal STP $\OptimumSol$ that minimizes $\STPEnergyTimeslot$ is given by
	\begin{align}\label{eq:FirstOptimumSol}
	\OptimumSol
	=
	\begin{cases}
	\SecondLocalOptimum, & \text{if } \FirstLocalOptimum<\CoverageSol<\SecondLocalOptimum,\\
	\CoverageSol, & \text{if } \SecondLocalOptimum\leq\CoverageSol,\\
	\underset{\SD \in \left\{\CoverageSol,\ \SecondLocalOptimum\right\}}{\arg\min} \mathcal{E}\left(\SD\right), & \text{if } \CoverageSol\leq\FirstLocalOptimum 
	\end{cases}
	\end{align}
	where $\FirstLocalOptimum$ and $\SecondLocalOptimum$ are two stationary points of $\STPEnergyTimeslot$, and $\FirstLocalOptimum<\SecondLocalOptimum$, and $\CoverageSol$ is given in \eqref{eq:MinimumCoverageSTP}.
	For $\Ni<9$ in the noise-limited environment and $\Ni<\FirstN$ in the interference-limited environment, we have
	\begin{align}
	\label{eq:SecondOptimumSol}
	\OptimumSol
	=
	\CoverageSol .
	\end{align}
\end{lemma}
\begin{IEEEproof}
	When there exist two stationary points, since $\lim\limits_{\SD\rightarrow1}\frac{d\mathcal{E}\left(\SD\right)}{d\SD}>0$ in \eqref{eq:FirstEnergyDerivative} and \eqref{eq:SecondEnergyDerivative}, we can know that $\FirstLocalOptimum$ and $\SecondLocalOptimum$ are local maximum and minimum, respectively.
	Since $\SD\geq\CoverageSol$ from the constraint in \eqref{eq:MinimumCoverageSTP}, when $\FirstLocalOptimum<\CoverageSol<\SecondLocalOptimum$, $\SecondLocalOptimum$ becomes the optimal $\SD$. When $\CoverageSol\geq\SecondLocalOptimum$, since $\STPEnergyTimeslot$ keeps increasing with $\SD$, $\CoverageSol$ becomes the optimal.
	On the other hand, when $\CoverageSol\le\FirstLocalOptimum$, the optimal \ac{STP} can be selected between  $\CoverageSol$ and $\SecondLocalOptimum$ as the one providing smaller $\STPEnergyTimeslot$.
	Lastly, when there is no stationary point, $\STPEnergyTimeslot$ is a strictly increasing function of $p_s$. Therefore, $\CoverageSol$ becomes the optimal \ac{STP}.
\end{IEEEproof}
Once we obtain the optimal STP $\OptimumSol$ as in Lemma~\ref{lem:optimal}, we can also present the optimal transmission power as
\begin{align}\label{eq:OptimumPowerSol}
\OptimumPowerSol
=
\STPinvfunctionOpt,
\end{align}
where $\STPinvfunction$ is given in \eqref{eq:STPinvSNR} and \eqref{eq:STPinvSIR}.
In addition, from Lemma~\ref{lem:optimal}, we can see the effect of $\Ni$ on the optimal transmission power $\TransPower^*$ as follows.
\begin{corollary} \label{col:optimal_power}
	The optimal transmission power $\OptimumPowerSol$ increases as the maximum number of retransmissions $\Ni$ increases.
\end{corollary}
\begin{IEEEproof}
	As $\Ni$ increases, the AoI violation probability in \eqref{eq:MinimumCoverageSTP} increases, which makes $\CoverageSol$ in \eqref{eq:MinimumCoverageSTP} larger. 
	We can also see that when there exist two stationary points, the increase of $\Ni$ makes the gap between $\FirstLocalOptimum$ and $\SecondLocalOptimum$ bigger, as also shown in Fig. \ref{fig:Energy_derivative}.
	From these results, $\OptimumSol$ in \eqref{eq:FirstOptimumSol} increases with $\Ni$ since both $\SecondLocalOptimum$ and $\CoverageSol$ increase.
	On the other hand, when the stationary point does not exist, as can be seen in \eqref{eq:SecondOptimumSol}, $\OptimumSol$ is only determined by $\CoverageSol$ which increases as $\Ni$ increases.
\end{IEEEproof}
From Corollary~\ref{col:optimal_power}, we can see that it is better to use larger transmission power $\TransPower$ as $\Ni$ increases for lower average energy consumption.
	\section{Numerical Results}\label{sec:numerical}
	\begin{table}[!t]
	\caption{Parameter values if not otherwise specified \label{table:parameter}} 
	\begin{center}
		\renewcommand{\arraystretch}{1.5}
		\begin{tabular}{l l | l l}
			\hline 
			{\bf Parameters} & {\bf Values} & {\bf Parameters} & {\hspace{0.32cm}}{\bf Values} \\ 
			\hline 
			\hspace{0.15cm}$\lambda_1$, $\lambda_2$ [nodes/$m^2$] & \hspace{0.2cm}$8.7\times10^{-5}$ 
			& \hspace{0.12cm}$\PL$ & \hspace{0.2cm}$3.5$ \\ \hline 
			\hspace{0.15cm}$P_1$ [mW] & \hspace{0.2cm}$40$ 
			& \hspace{0.12cm}$P_2$ [mW] & \hspace{0.2cm}$30$  \\ \hline 
			\hspace{0.15cm}$\distance$ [m] & \hspace{0.2cm}$20$ 
			& \hspace{0.12cm}$\delta$ & \hspace{0.2cm}$1$  \\ \hline 
			\hspace{0.15cm}$\TransTime$ & \hspace{0.2cm}$1$ 
			& \hspace{0.12cm}$\CommunicationTime$ & \hspace{0.2cm}$1$ \\ \hline 
			\hspace{0.15cm}$\CoverageThres$ & \hspace{0.2cm}$0.1$ 
			& \hspace{0.12cm}$\eta$ & \hspace{0.2cm}$0.6$ \\ \hline 
			\hspace{0.15cm}$\epsilon$ & \hspace{0.2cm}$0.6$
			& \hspace{0.12cm}$\Noise$ & \hspace{0.2cm}$10^{-5}$  \\ \hline 
			\hspace{0.15cm}$E_\text{s}$ [mW] & \hspace{0.2cm}$1$
			& \hspace{0.12cm}$ $ & \hspace{0.2cm}$ $  \\ 
			\hline
		\end{tabular}\vspace{-0.3cm}
	\end{center}
\end{table}%
	In this section, we present the numerical results of the AoI performance for the sensor monitoring system. 
	We have also verified the analysis results with the Monte-Carlo simulation, which is conducted by MATLAB.
	%
	Unless otherwise specified, the values of system parameters presented in Table \ref{table:parameter} are used.
\begin{figure}[t!]
	\begin{center}   
		{ 
			\psfrag{X}[bl][bl][0.7] {$\TransPower$ (dBm)}
			\psfrag{Y}[bl][bl][0.7] {Average AoI $\bar{\Delta}$}
			\psfrag{A2}[bl][bl][0.7] {$\Ni=10, \TransTime=1$}
			\psfrag{A3}[bl][bl][0.7] {$\Ni=5, \TransTime=2$}
			\psfrag{AAAAAAAAAAA}[bl][bl][0.7] {$\Ni=20, \TransTime=1$}
			\psfrag{A1}[bl][bl][0.7] {$\Ni=10, \TransTime=2$}
			\psfrag{A4}[bl][bl][0.7] {Simulation}
			%
			\includegraphics[width=1.00\columnwidth]{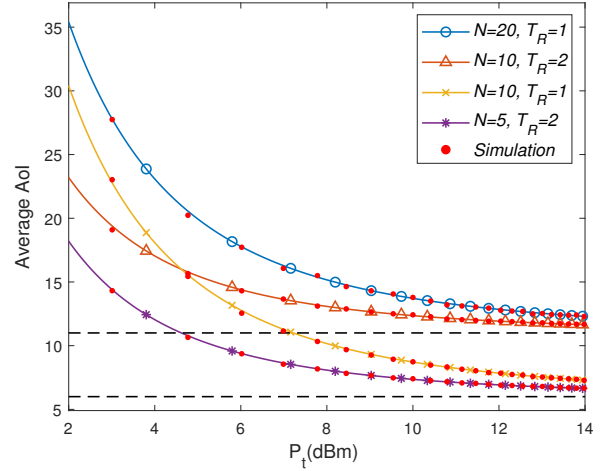}
		}
	\end{center}
	\caption{Average AoI as a function of transmission power $\TransPower$ for different values of $\Ni$ and $\TransTime$.}
	\vspace{0.7mm}
	\label{fig:AverageAoI}
\end{figure}

Figure \ref{fig:AverageAoI} presents the average AoI as a function of the transmission power $P_t$ with the different values of the maximum number of transmissions $\Ni$ and the transmission interval $\TransTime$. From Fig. \ref{fig:AverageAoI}, we can see that the analysis results fit well with the simulation results. Obviously, the average AoI decreases as $\TransPower$ increases. This is because the peak AoI of each update decreases due to the less retransmissions. 
We can also see that the average AoI increases as $\Ni$ increases because of the longer sensing period $\SamplePeriod$.
Furthermore, we can see that the average AoIs of two cases with $NT_R=10$ and $NT_R=20$ converge to the same value (i.e., $6$ and $11$, respectively) because large $\TransPower$ make the communication successful without retransmission and the updates can occur every sensing period $\SamplePeriod$.
\begin{figure}[t!]
	\begin{center}   
		{ 
			\psfrag{X}[bl][bl][0.7] {$\TransPower$ (dBm)}
			\psfrag{Y}[bl][bl][0.7] {AoI violation probability}
			\psfrag{A1}[bl][bl][0.7] {$\Ni=20, d=25$}
			\psfrag{A2}[bl][bl][0.7] {$\Ni=20, d=20$}
			\psfrag{A3}[bl][bl][0.7] {$\Ni=15, d=20$}
			\psfrag{A4}[bl][bl][0.7] {$\Ni=10, d=20$}
			\psfrag{A5}[bl][bl][0.7] {$\Ni=5, d=20$}
			\psfrag{AAAAAAAAAAA}[bl][bl][0.7] {Simulation}
			%
			\includegraphics[width=1.00\columnwidth]{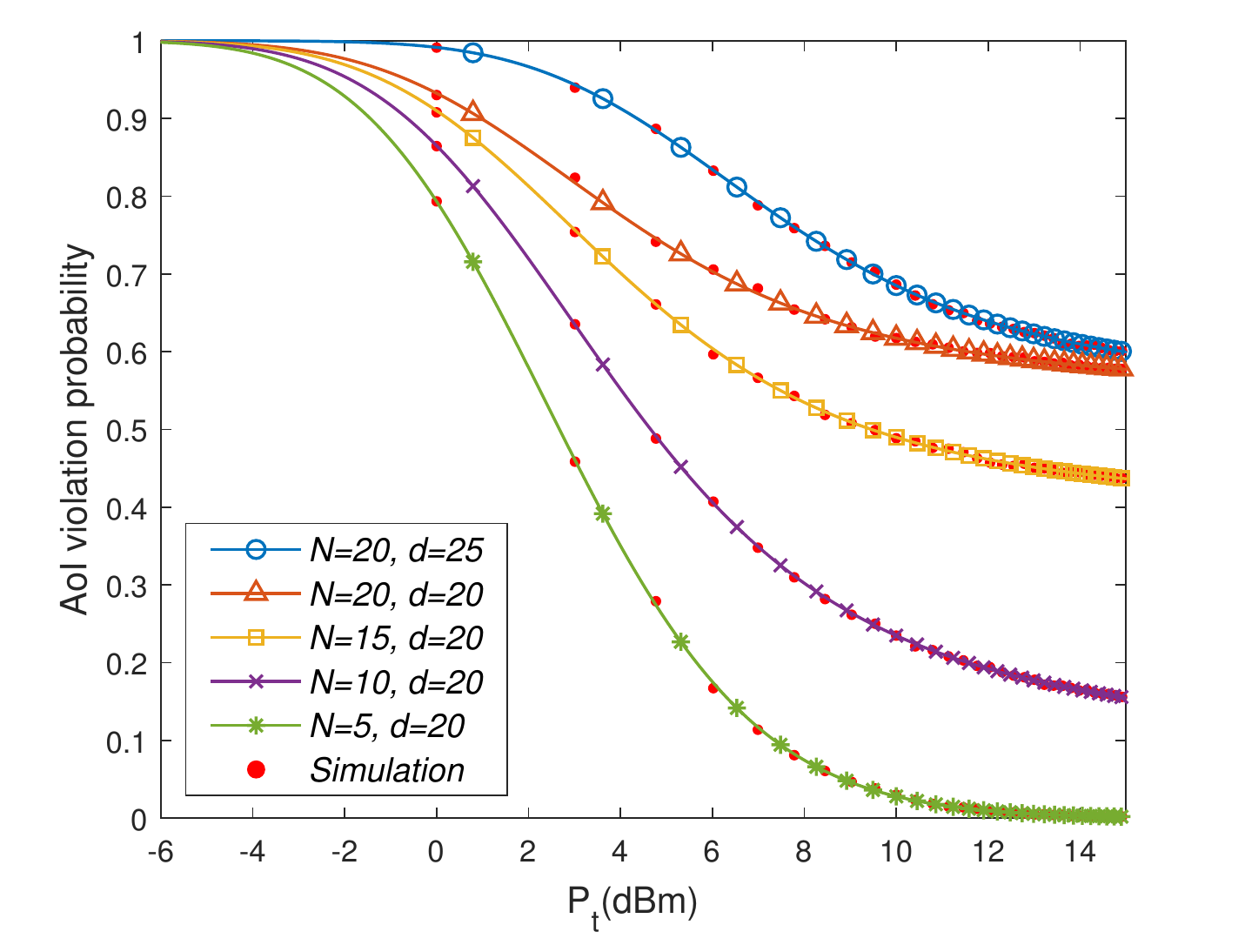}
			\vspace{-8mm}
		}
	\end{center}
	\caption{AoI violation probability $\PV{\Viol}$ as a function of the transmission power $\TransPower$ with $\Viol=10$ for different values of $\Ni$.}
	\vspace{-2mm}
	\label{fig:AoIviolationprobability}
\end{figure}

Figure \ref{fig:AoIviolationprobability} shows the AoI violation probability as a function of transmission power $\TransPower$ for different values of $\Ni$ with $\Viol=10$. As discussed in Corollary 1, the AoI violation probability decreases as $\TransPower$ increases due to the smaller number of retransmissions. 
We can see that the AoI violation probability increases as $\Ni$ increases because of the longer sensing period $\SamplePeriod$.
We can also see that as the main link distance increases, the AoI violation probability increases because of low \ac{STP}.

\begin{figure}[t!]
	\begin{center}   
		{ 
			\psfrag{AAAAAA}[bl][bl][0.7] {$\Ni=9$}
			\psfrag{A1}[bl][bl][0.7] {$\Ni=10$}
			\psfrag{A2}[bl][bl][0.7] {$\Ni=11$}
			\psfrag{X}[bl][bl][0.7] {$\TransPower \left(\text{dBm}\right)$}
			\psfrag{Y}[bl][bl][0.7] {Average energy consumption $\EnergyTimeslot$}
			\includegraphics[width=1.00\columnwidth]{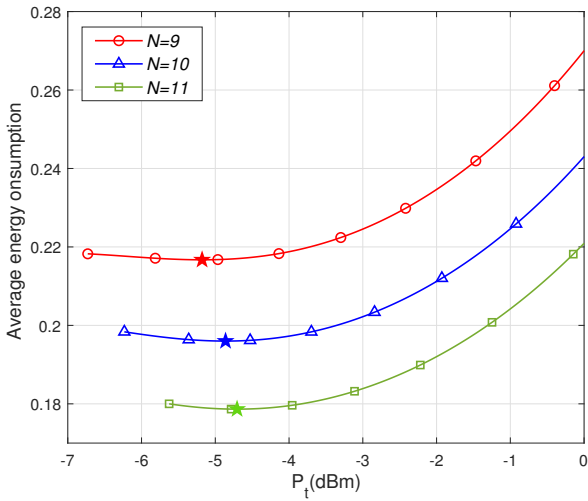}
		}
	\end{center}
	\caption{Average energy consumption as a function of transmission power $\TransPower$ with $\TransTime=1$ for different values of $\Ni$ in noise-limited environment.}
	\vspace{0.7mm}
	\label{fig:Energy_SNR}
\end{figure}
     Figure \ref{fig:Energy_SNR} presents the average energy consumption as a function of $\TransPower$ for different values of $\Ni$ in the noise-limited environment. 
	Here, we use $u=1.76\times{10}^{-2}$, $v=1.2\times{10}^{-3}$, and $\MaximumRadius=3$. 
	Each graph is plotted for a feasible transmission power which satisfies the coverage constraint \eqref{eq:MinimumCoverageSTP}. As can be seen in Fig.~\ref{fig:Energy_SNR}, $\EnergyTimeslot$ decreases and then increases. 
	This is because, for small $\TransPower$, as $\TransPower$ increases, the reduced number of retransmissions (which lowers $\EnergyTimeslot$) more dominantly affect than the increased $\TransPower$ (which enlarges $\EnergyTimeslot$). 
	However, when $\TransPower$ is larger than a certain value,  
	the retransmission number is not reduced anymore as the STP is almost one, so increasing $\TransPower$ increases $\EnergyTimeslot$. 
	Therefore, there exist the optimal transmission power $\TransPower^*$ that minimizes the average energy consumption
	such as $\TransPower^*=-5.18$  dBm for $\Ni=9$, which is consistent with the first case result of Lemma~2.
	Moreover, as discussed in Corollary~\ref{col:optimal_power}, we can see that the optimal transmission power $\TransPower^*$ increases with the maximum number of transmissions $\Ni$, which indicates that the sensor needs to use larger transmission power when the sampling period becomes longer.

\begin{figure}[t!]
	\begin{center}   
		{ 
			\psfrag{AAAAAAAAAAAAAAAAAAAAAAAAAAAA}[bl][bl][0.7] {$\Ni=20$, general environment}
			\psfrag{A1}[bl][bl][0.7] {$\Ni=25$, general environment}
			\psfrag{A2}[bl][bl][0.7] {$\Ni=30$, general environment}
			\psfrag{A3}[bl][bl][0.7] {$\Ni=20$, interference-limited assumption}
			\psfrag{A4}[bl][bl][0.7] {$\Ni=25$, interference-limited assumption}
			\psfrag{A5}[bl][bl][0.7] {$\Ni=30$, interference-limited assumption}
			\psfrag{X}[bl][bl][0.7] {$\TransPower \left(\text{dBm}\right)$}
			\psfrag{Y}[bl][bl][0.7] {Average energy consumption $\EnergyTimeslot$}
			\includegraphics[width=1.00\columnwidth]{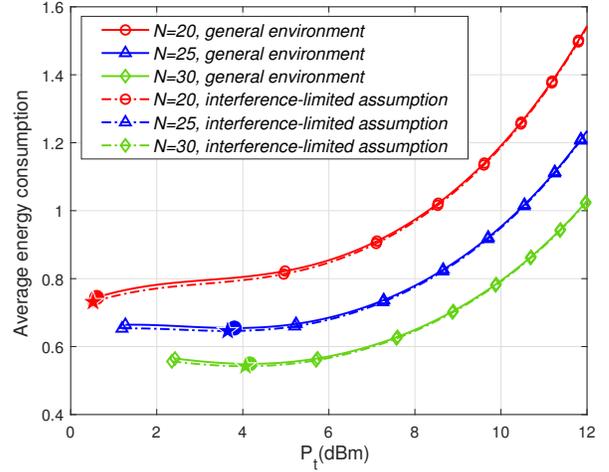}
			\vspace{-8mm}
		}
	\end{center}
	\caption{Average energy consumption as a function of the transmission power $\TransPower$ for different values of $\Ni$ {for general environment and with interference-limited assumption when $\Noise=10^{-6}$.}}
	\vspace{-2mm}
	\label{fig:Energy_SIR}
\end{figure}

Figure \ref{fig:Energy_SIR} presents the average energy consumption as a function of $\TransPower$ for different $\Ni$ with and without the interference-limited environment assumption. For both cases, we use $u=1.76\times{10}^{-2}$, $v=4.4\times{10}^{-4}$, and $\MaximumRadius=3$. In Fig. \ref{fig:Energy_SIR}, circle symbols represent the optimal transmission power $\OptimumPowerSol$ that minimizes the average energy consumption in general environment, and star symbols represent $\OptimumPowerSol$ with the interference-limited assumption.
Since the STP is lower in the general environment than that with the interference-limited assumption for the same $\TransPower$, the average energy consumption in the general environment is higher than that with the interference-limited assumption.
Similar to the results in the noise-limited environment, 
we can see that $\TransPower^*$ is higher for larger $\Ni$. 
Since the STP is higher in the case with the interference-limited assumption than the general environment 
as the noise power is ignored, 
we can observe that 
$\TransPower^*$ as well as the average energy consumption are higher in the general environment 
than those with the interference-limited assumption. 
Therefore, $\TransPower^*$ with the interference-limited assumption can provide the lower bound of 
the real optimal transmission power, 
but the results with the interference-limited assumption become more valid and similar to the real optimal value as the effect of the interference increases (i.e., as $\lambda_{\text{I}}$ increases).


\begin{figure}[t!]
	\begin{center}   
		{ 
			\psfrag{X}[bl][bl][0.7] {$\TransPower \left(\text{dBm}\right)$}
			\psfrag{Y}[bl][bl][0.7] {Maximum number of retransmission $\Ni$}
			\includegraphics[width=1.00\columnwidth]{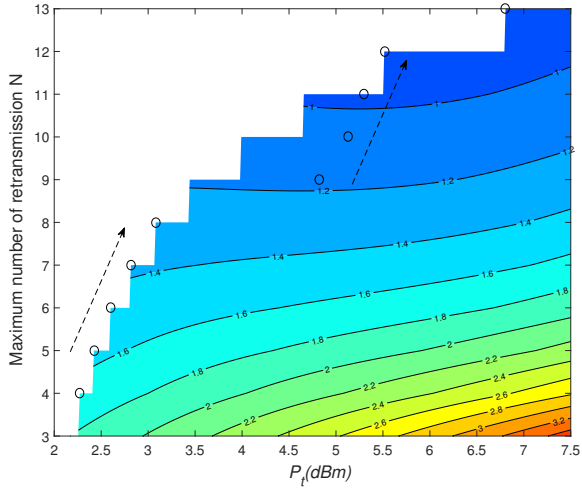}
		}
	\end{center}
	\caption{Average energy consumption as functions of the transmission power $\TransPower$ and the maximum number of transmissions $\Ni$ with $\TransTime=1$ in noise-limited case. The optimal transmission powers that minimize the average energy consumption are marked by circles.}
	\label{fig:Energy_3D}
\end{figure}

Figure \ref{fig:Energy_3D} shows the average energy consumption as functions of $\TransPower$ and $\Ni$ in the noise-limited environment. The graph is plotted for the feasible range of $\TransPower$, which satisfies the coverage constraint in \eqref{eq:MinimumCoverageSTP}.
In addition, the circle symbols in Fig. \ref{fig:Energy_3D} refer to the optimal transmission power $\OptimumPowerSol$ which minimizes the average energy consumption for each $\Ni$.
From Fig.~\ref{fig:Energy_3D}, we can see that the minimum transmission power $P_\text{cov}$ which satisfies the coverage constraint increases as $\Ni$ increases. It is because $p_\text{cov}$ that satisfies the target violation threshold in \eqref{eq:MinimumCoverageSTP} increases as the AoI violation probability increases for each $\SD$ as $\Ni$ increases. We can also see that when $\Ni\geq9$, there exist the stationary points as discussed in Proposition~1, so $\OptimumPowerSol$ becomes higher than $P_\text{cov}$. 
As also shown in Corollary~2, we can also observe that $\OptimumPowerSol$ increases as $\Ni$ increases.


\section{Conclusion}
In this paper, we consider the sensor monitoring system, which requires maintaining fresh data at the \ac{DC}.   
After defining the \ac{ETS} coverage by reflecting the spatial-temporal correlation of the sensing information, 
we derive the average AoI and the AoI violation probability. 
We then finally provide the $\eta$-coverage probability, and analytically show it increases with the transmission power of the sensor. 
%
We also provide the optimal transmission power of the sensor that minimizes the average energy consumption of the sensor. 
%
Our results provide some insights on the energy-efficient design of the sensor monitoring system that requires securing fresh data. 
Specifically, when the maximum number of retransmissions is small, the minimum power that satisfies the constraint of the $\eta$-coverage probability becomes the optimal. 
The optimal transmission power also increases with the maximum number of retransmissions 
%
%
This work paves the way to efficient design of the AoI-sensitive sensor networks. 
\begin{appendix}\label{App:thm1}
	\subsection{Proof of Theorem~\ref{thr:APP}}
	We first analyze the violated time $\VL{k}$, which is defined as
	\begin{align}\label{eq:gk}
	\VL{k} \hspace{-0.7mm} = \hspace{-0.7mm}
	\begin{cases}
	0, & \mbox{if} \,\,\, \Viol>\SamplePeriod+\SuccTrans{k},\\
	\SamplePeriod \hspace{-0.7mm} + \hspace{-0.7mm}\SuccTrans{k} \hspace{-0.7mm} - \hspace{-0.7mm}\Viol, & \mbox{if} \,\,\, \RealSuccTrans{k-1}\leq\Viol\leq \SamplePeriod \hspace{-0.7mm}+ \hspace{-0.7mm}\SuccTrans{k},\\
	\WaitTime{k-1} \hspace{-0.7mm}+ \hspace{-0.7mm}\SuccTrans{k}, & \mbox{if} \,\,\, \Viol<\RealSuccTrans{k-1},
	\end{cases}
	\end{align}
	where $\SamplePeriod+\SuccTrans{k}$ represents the $k$th peak of AoI. Note that $\RealSuccTrans{k-1}\leq\SamplePeriod, \forall k$.
	From \eqref{eq:gk}, by using the independence of $\RealSuccTrans{k-1}$ and $\SuccTrans{k}$\footnote{Note that $\SuccTrans{k}$ is a dependent variable on $\RealSuccTrans{k}$. However, since $\RealSuccTrans{k-1}$ and $\RealSuccTrans{k}$ are independent, $\SuccTrans{k}$ and $\RealSuccTrans{k-1}$ are independent as well.}, the expectation of $\VL{k}$ is given by 
	%
	\begin{align}\label{eq:ViolationLength}
	\nonumber
	\Expt{\VL{k}}
	=&\ 
	\FirstCondEx \Prob{\RealSuccTrans{k-1}\leq\Viol}\Prob{\SamplePeriod+\SuccTrans{k}>\Viol}
	\\
	&+\SecondCondEx \Prob{\RealSuccTrans{k-1}>\Viol}
	\end{align}
	where $\FirstCondEx$ and $\SecondCondEx$ are
	\begin{align}\label{eq:ConditionalExp}
	\nonumber
	\FirstCondEx
	&=
	\ConExpt{\SamplePeriod+\SuccTrans{k}-\Viol\hspace{-0.5mm}}{\hspace{-0.5mm}\RealSuccTrans{k-1}\hspace{-0.5mm}\leq\hspace{-0.5mm}\Viol,\SamplePeriod+\SuccTrans{k}\hspace{-0.5mm}>\hspace{-0.5mm}\Viol}
	\\
	\SecondCondEx
	&=
	\ConExpt{\WaitTime{k-1}+\SuccTrans{k}}{\RealSuccTrans{k-1}>\Viol}.
	\end{align}
	Here, $\Expt{\VL{k}}$ needs to be analyzed differently for the cases of $\Viol\leq\SamplePeriod$ and $\Viol>\SamplePeriod$ as follows.
	\subsubsection{Case 1 ($\Viol\leq\Ni\TransTime$)}
	For $\Viol\leq\Ni\TransTime$, we have
	\begin{align}
	&\Prob{\RealSuccTrans{k-1}\hspace{-0.5mm}\leq\Viol}\label{eq:secondtermofprobability}
	\hspace{-0.7mm}
	=
	\hspace{-0.7mm}
	\sum_{n=1}^{\kai}\frac{{\left(1\hspace{-0.7mm}-\hspace{-0.7mm}\SD\right)}^{n-1}\SD}{1\hspace{-0.7mm}-\hspace{-0.7mm}\left(1\hspace{-0.7mm}-\hspace{-0.7mm}\SD\right)^N}
	=
	\frac{1\hspace{-0.7mm}-\hspace{-0.7mm}\left(1\hspace{-0.7mm}-\hspace{-0.7mm}\SD\right)^{\kai}}{1\hspace{-0.7mm}-\hspace{-0.7mm}\left(\hspace{-0.5mm}1\hspace{-0.7mm}-\hspace{-0.7mm}\SD\right)^N},
	\\
	&\Prob{\RealSuccTrans{k-1}>\Viol}=1-\Prob{\RealSuccTrans{k-1}\leq\Viol}
	=
	\frac{\left(\hspace{-0.2mm}1\hspace{-0.7mm}-\hspace{-0.7mm}\SD\hspace{-0.2mm}\right)^{\kai}\hspace{-0.7mm}-\hspace{-0.7mm} \left(\hspace{-0.2mm}1\hspace{-0.7mm}-\hspace{-0.7mm}\SD\hspace{-0.2mm}\right)^{\hspace{-0.3mm}\Ni}}{1\hspace{-0.7mm}-\hspace{-0.7mm}\left(1\hspace{-0.7mm}-\hspace{-0.7mm}\SD\right)^\Ni}, 
	\label{eq:secondsecondtermofprobability}
	\\
	&\Prob{\SamplePeriod+\SuccTrans{k}>\Viol}=1,
	\end{align}
	{where $\kai$ is the maximum number of transmissions, allowed within $\Viol$, given by}
	\begin{align}
		\kai=\NV.\label{eq:maximumnumoftransmissions}
	\end{align}
	Similar to \eqref{eq:ExpectationX}, $\FirstCondEx$ can be obtained as
	\begin{align}\label{eq:firsttermofexpectation}
	\FirstCondEx
	=
	\SamplePeriod-\Viol+\left(\frac{1}{\SD}-1\right)\TransTime+\CommunicationTime.
	\end{align}
	%
	%
	%
	%
	%
	%
	%
	In \eqref{eq:ConditionalExp}, $\SecondCondEx$ can also be obtained as
	\begin{align}\label{eq:thirdtermofexpectation}
	\SecondCondEx&=
	\ConExpt{\SamplePeriod-\RealSuccTrans{k-1}+\SuccTrans{k}}{\RealSuccTrans{k-1}>\Viol}
	\nonumber
	\\
	&=
	\SamplePeriod+\frac{\TransTime}{\SD}-\sum_{n=\kai+1}^{N}\frac{\SD\TransTime n\left(1-\SD\right)^{n-1}}{\CP^{\kai}-\left(1-\SD\right)^{\Ni}}
	\nonumber \\
	&=
	\SamplePeriod+\frac{\TransTime}{\SD}
	\nonumber\\
	& - \frac{\TransTime \hspace{-0.5mm} \left\{ \hspace{-0.8mm}
		\left(\hspace{-0.2mm}1\hspace{-0.7mm}-\hspace{-0.7mm}\SD\hspace{-0.3mm}\right)^{\kai} \hspace{-0.8mm} \left(\hspace{-0.5mm}1\hspace{-0.5mm} + \hspace{-0.5mm} \SD \kai \hspace{-0.3mm}\right)\hspace{-0.7mm} - \hspace{-0.7mm} \left(\hspace{-0.2mm}1\hspace{-0.7mm}-\hspace{-0.7mm}\SD\hspace{-0.3mm}\right)^{\Ni} \hspace{-0.5mm} \left(\hspace{-0.2mm}1\hspace{-0.5mm} + \hspace{-0.5mm} {\SD}N \hspace{-0.3mm}\right) \hspace{-0.5mm}\right\}} {\SD\left\{\CP^{\kai}\hspace{-0.3mm} - \hspace{-0.3mm} \left(1\hspace{-0.3mm} - \hspace{-0.3mm}\SD\right)^{\Ni}\right\}}. \hspace{-1mm}
	\end{align}
	By substituting \eqref{eq:secondtermofprobability}$-$\eqref{eq:thirdtermofexpectation} into \eqref{eq:ViolationLength}, we can express $\Expt{\VL{k}}$ as
	\begin{align}\label{eq:ViolationFirst}
	\Expt{\VL{k}}
	&=
	\frac{\left\{\SD\left(\CommunicationTime-\TransTime-\Viol\right)+\TransTime\right\}\left\{1-{\CP}^{\kai}\right\}}{\SD\left\{\CPP\right\}}
	\nonumber
	\\
	&\quad+\frac{\TransTime\left\{\Ni-\kai{\CP}^{\kai}\right\}}{\CPP}.
	\end{align}
	Finally, by dividing \eqref{eq:ViolationFirst} by \eqref{eq:Interarrival} and replacing $\kai$ with \eqref{eq:maximumnumoftransmissions}, $\PV{\Viol}$ for the case of $\Viol\leq\SamplePeriod$ is presented as the first equation in \eqref{eq:SecondViolationProbability}.
	%
	
	\subsubsection{Case 2 ($\Viol>\SamplePeriod$)}
	In \eqref{eq:ViolationLength}, we have 
	\begin{align}\label{eq:secondfirstprobability}
	\Prob{\RealSuccTrans{k-1}\leq\Viol}
	&\overset{\underset{\mathrm{(a)}}{}}{=}
	1, \quad
	\Prob{\RealSuccTrans{k-1}>\Viol}
	\overset{\underset{\mathrm{(a)}}{}}{=}
	0
	\\
	\Prob{\SamplePeriod+\SuccTrans{k}>\Viol}
	&=
	\sum_{n=\kai-\Ni+1}^{\infty}\hspace{-1mm}\CP^{n-1}\SD
	\nonumber
	\\
	&=\CP^{\kai-\Ni},
	\end{align}
	where (a) is obtained by the fact $Z_{k-1} \le NT_R$.
	We also obtain $\FirstCondEx$ as
	\begin{align}\label{eq:secondfirstconex}
	\FirstCondEx
	=
	\TransTime\left(\frac{1}{\SD}+\kai-1\right)+\CommunicationTime-\Viol.
	\end{align}
	Here, $\SecondCondEx$ is not presented since $\Prob{\RealSuccTrans{k-1}>\Viol}=0$ in \eqref{eq:ViolationLength}.
	By substituting \eqref{eq:secondfirstprobability}$-$\eqref{eq:secondfirstconex} into \eqref{eq:ViolationLength}, we finally obtain $\Expt{\VL{k}}$ for the case of $\Viol>\SamplePeriod$ as
	\begin{align}\label{eq:Secondgk}
	\Expt{\VL{k}}
	\nonumber
	&=
	\left\{\TransTime\left(\frac{1}{\SD}+\kai-1\right)+\CommunicationTime-\Viol\right\}
	\\
	&\times\CP^{\kai-\Ni}.
	\end{align}
	By dividing \eqref{eq:Secondgk} by \eqref{eq:Interarrival} and replacing $\kai$ with \eqref{eq:maximumnumoftransmissions}, we obtain $\PV{\Viol}$ as the second equation in \eqref{eq:SecondViolationProbability}.
\end{appendix}
	\bibliographystyle{IEEEtran}
	\bibliography{Bibtex/StringDefinitions,Bibtex/IEEEabrv,Bibtex/ISCGroup}
\end{document}

%% file: SupportDocuments/acronym.tex
\acrodef{PAoI}{peak age of information}
\acrodef{WSN}{wireless sensor network}

\acrodef{DC}{data center}
\acrodef{MEC}{mobile edge computing}

\acrodef{AWGN}{additive white Gaussian noise}
\acrodef{PPP}{Poisson point process}
\acrodef{HPPP}{homogeneous Poisson point process}
\acrodefplural{PPP}[PPPs]{Poisson point processes}

\acrodef{PDF}{probability density function}
\acrodef{PMF}{probability mass function}
\acrodef{CDF}{cumulative distribution function}
\acrodef{CCDF}{complementary cumulative distribution function}
\acrodef{SIR}{signal-to-interference ratio}
\acrodef{SINR}{signal-to-interference-plus-noise ratio}
\acrodef{SNR}{signal-to-noise ratio}

\acrodef{STP}{successful transmission probability}
\acrodef{ETS}{error-tolerable sensing}

\acrodef{P-K}{Pollaczek-Khinchin}
\acrodef{KKT}{Karush-Kuhn-Tucker}
\acrodef{IoT}{Internet of Things}
\acrodef{URLLC}{ultra reliable low latency communication}
\acrodef{AoI}{age of information}
\acrodef{ACK}{acknowledgement}
\acrodef{NACK}{negative acknowledgement}

\acrodef{i.i.d.}{independent, identically distributed}

\acrodef{HARQ}{hybrid automatic repeat request}
\acrodef{ARQ}{automatic repeat-request}
\acrodef{MTC}{machine type communications}
\acrodef{IoT}{Internet of Things}

%% file: SupportDocuments/defmetric.tex
\usepackage{color}
\usepackage{dsfont}
\usepackage{bbm}

\newtheorem{theorem}{Theorem}
\newtheorem{lemma}{Lemma}
\newtheorem{corollary}{Corollary}
\newtheorem{proposition}{Proposition}

\newtheorem{problem}{Problem}

\newcommand{\be}{\begin{enumerate}}
	\newcommand{\ee}{\end{enumerate}}
\newcommand{\bi}{\begin{itemize}}
\newcommand{\ei}{\end{itemize}}
\newcommand{\Prob}[1]{\mathbb{P}\left[#1\right]}

\newcommand{\Expt}[1]{\mathbb{E}\left[#1\right]}

\newcommand{\ConExpt}[2]{\mathbb{E}\left[#1\mid#2\right]}

\newcommand{\Interference}{I}
\newcommand{\InterferenceDensity}{\lambda_{j}}

\newcommand{\InterferenceGamma}{C_{1}\left(\alpha\right)}

\newcommand{\STPfunction}{f_{\text{s}}\left(\TransPower\right)}
\newcommand{\STPinvfunction}{f^{-1}_{\text{s}}\left(\SD\right)}
\newcommand{\STPinvfunctionOpt}{f^{-1}_{\text{s}}\left(\SD^*\right)}

\newcommand{\TransTime}{T_{\text{R}}}
\newcommand{\SD}{p_\text{s}}
\newcommand{\SamplePeriod}{NT_\text{R}}
\newcommand{\Ni}{N}
\newcommand{\CommunicationTime}{T_{\text{o}}}

\newcommand{\Cov}{\rho\left(r, t\right)}
\newcommand{\Error}{\epsilon\left(r, t\right)}

\newcommand{\Sensinginstant}[1]{t_{{#1}}}
\newcommand{\Arrivalinstant}[1]{\hat{t}_{{#1}}}
\newcommand{\RealSensinginstant}[1]{\bar{t}_{{#1}}}
\newcommand{\TD}{\tau}
\newcommand{\numberofupdates}{L\left(T\right)}
\newcommand{\AoI}{\Delta\left(t\right)}
\newcommand{\AAoI}{\bar{\Delta}}

\newcommand{\SuccTrans}[1]{X_{#1}}
\newcommand{\RealSuccTrans}[1]{Z_{#1}}
\newcommand{\WaitTime}[1]{Y_{#1}}

\newcommand{\InterArea}[1]{Q_{#1}}

\newcommand{\CoverageThres}{\theta_{\text{th}}}

\newcommand{\RadSquare}{r^2\left(\AoI, \CoverageThres\right)}

\newcommand{\SensorCoverage}{S_{\text{c}}\left(\CoverageThres\right)}
\newcommand{\MaximumRadius}{r_{\text{max}}}
\newcommand{\CoverageSol}{p_{\text{cov}}}

\newcommand{\Etacoverage}{\mathcal{P}_{\text{c}}\left(\eta\right)}

\newcommand{\Correlation}{\rho\left(l, \TD\right)}
\newcommand{\Radius}{r\left(\tau, \CoverageThres\right)}
\newcommand{\kai}{\chi}

\newcommand{\FirstCondEx}{\mathcal{P}_{g, 1}}
\newcommand{\SecondCondEx}{\mathcal{P}_{g, 2}}

\newcommand{\PV}[1]{\mathcal{P}_{\text{v}}\left({#1}\right)}
\newcommand{\PVinv}[1]{\mathcal{P}^{-1}_{\text{v}}\left({#1}\right)}
\newcommand{\VL}[1]{g_{#1}}
\newcommand{\IL}[1]{U_{#1}}
\newcommand{\Viol}{v_\text{th}}

\newcommand{\CP}{\left(1-\SD\right)}
\newcommand{\CPP}{1-\left(1-\SD\right)^{\Ni}}

\newcommand{\NV}{\left\lfloor\hspace{-0.5mm}{\frac{\Viol-\CommunicationTime}{\TransTime}}\hspace{-0.5mm}\right\rfloor\hspace{-0.3mm}+\hspace{-0.3mm}1} 
\newcommand{\NVN}{\left\lfloor\hspace{-0.5mm}{\frac{\Viol-\CommunicationTime}{\TransTime}}\hspace{-0.5mm}\right\rfloor-\Ni} 
\newcommand{\FirstNominator}[1]{{\bar{N}}_{\text{v}, 1}\left({#1}\right)}

\newcommand{\EnergyTimeslot}{\mathcal{E}\left(\TransPower\right)}
\newcommand{\STPEnergyTimeslot}{\mathcal{E}\left(\SD\right)}

\newcommand{\Es}{E_\text{s}}
\newcommand{\FirstLocalOptimum}{s_{1}}
\newcommand{\SecondLocalOptimum}{s_{2}}
\newcommand{\EnergyDerivativeSNR}{f_{1}\left(\SD, \Ni\right)}
\newcommand{\EnergyDerivativeSIR}{f_2\left(\SD, \Ni, \PL\right)}

\newcommand{\FirstN}{{f}_{\text{I}}\left(\PL\right)}

\newcommand{\OptimumSol}{p_\text{s}^*}
\newcommand{\OptimumPowerSol}{P_\text{t}^*}

\newcommand{\arep}{\alpha_1}
\newcommand{\brep}{\beta_1}

\newcommand{\drep}{\beta_2}

\newcommand{\frep}{\gamma}

\newcommand{\CH}{h}
\newcommand{\PL}{\alpha}
\newcommand{\Noise}{N_\text{o}}
\newcommand{\distance}{d}
\newcommand{\TransPower}{P_\text{t}}

\newcommand{\NoiseCoefficient}{\xi}
\newcommand{\InterferenceCoefficient}{\zeta}